\documentclass[%
 aps,prd,reprint,
 amsmath,amssymb,
 nofootinbib
]{revtex4-2}

\usepackage[dvipsnames]{xcolor}
\usepackage{graphicx}
\usepackage{bm}
\usepackage{braket}
\usepackage{hyperref}
\hypersetup{colorlinks, urlcolor=BlueViolet, citecolor=Plum, linkcolor=PineGreen}
\usepackage{slashed}

\begin{document}

\title{A new idea for relating the asymmetric dark matter mass scale to the proton mass}

\author{Peter Cox}
 \email{peter.cox@unimelb.edu.au}
\author{Rafael E. Pérez}
 \email{rperezpenara@student.unimelb.edu.au}
\author{Raymond R. Volkas}
 \email{raymondv@unimelb.edu.au}
\affiliation{ARC Centre of Excellence for Dark Matter Particle Physics,\\
 School of Physics, The University of Melbourne, Victoria 3010, Australia}

\begin{abstract}
Asymmetric dark matter is a well-motivated approach to explain the apparent coincidence between the relic densities of visible and dark matter, $\Omega_D \simeq 5.4\Omega_b$. A complete explanation requires two components, a relation between the particle masses of the dark and visible matter, and a second relation between the number densities in each sector. In this work, we propose a new mechanism to address the former. We consider an extended $SU(3)_1 \times SU(3)_2$ colour group in the visible sector, with QCD embedded as the diagonal subgroup. A $\mathbb{Z}_2$ exchange symmetry then relates $SU(3)_2$ to a dark, confining $SU(3)_D$ sector. The dark matter is a composite state of dark fermions transforming in the fundamental representation of $SU(3)_D$. The spontaneously broken $\mathbb{Z}_2$ symmetry ultimately leads to a relation between the QCD and dark gauge couplings which, for suitable field content, gives rise to confinement scales of the same order of magnitude. The mechanism leads to a rich particle spectrum above the TeV scale which could be probed at future experiments. The model also naturally includes an axion solution to the strong CP problem.
\end{abstract}

\maketitle


\section{\label{sec:intro}Introduction}

Identifying the particle nature of dark matter (DM) is one of the most important goals in fundamental physics.\footnote{We acknowledge that asteroid-mass-scale primordial black holes are currently a viable DM candidate, but in this paper we consider particle DM only.} Because DM has been detected solely from gravitational interactions, at the present time we have relatively scant information about its properties: its non-gravitational couplings with ordinary matter are weak, it was non-relativistic at matter-radiation equality when cosmological structure started to form, it manifests as collisionless, and its cosmological mass density is
\begin{equation}
    \Omega_D \simeq 5.4\, \Omega_b \simeq 0.264,
    \label{eq:cosmocoincidence}
\end{equation}
where $\Omega$ denotes the ratio of the mass density of a species with the critical density ($D$ stands for DM, and $b$ for ordinary or baryonic matter). The first three properties give rise to the familiar collisionless cold dark matter paradigm.

In this paper, we analyse a new idea motivated by the fourth property. That measurement exhibits an apparent cosmological coincidence between the mass densities of dark and ordinary matter. Without a deep connection between the physics of how the ordinary and dark matter relic densities are generated, the similar mass densities must be taken as a genuine coincidence, and it is routine to do so. A simple example is provided by arguably the two favoured mechanisms: the baryon density arises from thermal leptogenesis in the type 1 seesaw model of neutrino masses, while DM is a thermal WIMP (weakly interacting massive particle) produced through freeze out. These mechanisms are unconnected, so  $\Omega_D \simeq 5.4\, \Omega_b$ has to arise as an accident. (The fact that $\Omega_D \sim 1$ may be explained by the ``WIMP miracle'', however.) If DM is instead comprised of axions generated through the misalignment mechanism, be it pre- or post-inflationary, the same conclusion holds.\footnote{An approach to the coincidence problem using axion DM in a cosmological relaxation scenario has recently been proposed~\cite{Banerjee:2024xhn,Brzeminski:2023wza}.}

The hypothesis of asymmetric dark matter (ADM)~\cite{Petraki:2013wwa,Zurek:2013wia} provides an approach to using $\Omega_D \simeq 5.4\, \Omega_b$ as a vital piece of information for uncovering the particle nature of DM. The essence is the simple postulate that since the baryon density originates from the cosmological baryon asymmetry, the DM density similarly arises from a DM particle-antiparticle asymmetry, with the two asymmetries being related through some high-temperature processes in the early universe. 

But there are two aspects to this relationship. A mass density is the product of a number density and a particle mass. In ADM models, one typically engineers similar number densities for baryons and DM, and then uses $\Omega_D \simeq 5.4\, \Omega_b$ to deduce the dark matter mass. The result, of course, is that this mass is an order-one number multiplied by the proton mass, so in the few GeV regime (typically). This common strategy fails, however, to really explain the apparent coincidence. Instead it shifts the coincidence to the question of why the DM mass is of a similar order as the proton mass.

A fully satisfactory ADM explanation of $\Omega_D \simeq 5.4\, \Omega_b$ thus requires a good rationale for a few-GeV DM mass scale. Since the proton mass is almost entirely due to QCD confinement energy, one is naturally drawn to DM being a dark baryon of a dark QCD sector whose confinement scale $\Lambda_D$ is of a similar order to the QCD confinement scale $\Lambda_\textrm{\tiny{QCD}}$, and preferably slightly larger.

Frameworks where the confinement scales can be related include: (a) mirror matter, with the mirror symmetry either exact~\cite{Foot:1991bp,Foot:1991py,Hodges:1993yb,Berezhiani:2000gw,Ignatiev:2003js,Foot:2003jt,Foot:2004wz,Foot:2004pq,Foot:2004pa,Das:2024tmx} or somewhat broken~\cite{Berezhiani:1996sz,An:2009vq,Cui:2011wk,Lonsdale:2014yua,Farina:2015uea,GarciaGarcia:2015pnn,Farina:2016ndq,Lonsdale:2017mzg,Lonsdale:2018xwd,Ibe:2019ena,Beauchesne:2020mih,Feng:2020urb,Ritter:2021hgu,Ibe:2021gil,Bodas:2024idn}, (b) evolution to an infrared fixed point in the QCD and dark QCD running couplings~\cite{Bai:2013xga,Newstead:2014jva,Ritter:2022opo,Ritter:2024sqv}, (c) ordinary and dark QCD originating from a parent unification group~\cite{Lonsdale:2014wwa,Lonsdale:2014yua,Murgui:2021eqf,Chung:2024nnj}, and (d) from the QCD vacuum~\cite{Chung:2024ezq}. The purpose of this paper is to propose a new framework for achieving related confinement scales. 

The gist of the idea is rather simple and uses a $\mathbb{Z}_2$ exchange symmetry to ultimately relate the confinement scales of QCD and a dark $SU(3)_D$. While this starting point is similar to mirror matter models, here, the $\mathbb{Z}_2$ symmetry does \emph{not} act on the SM fermions. Instead, we introduce an extended gauge group $SU(3)_1 \times SU(3)_2 \times SU(3)_D$, with the SM fermions charged only under $SU(3)_1$ and the $SU(3)_2$ and $SU(3)_D$ sectors related by the exchange symmetry. QCD is identified as the diagonal subgroup of $SU(3)_1 \times SU(3)_2$, with the spontaneous breaking $SU(3)_1 \times SU(3)_2 \to SU(3)_c$ occurring at a scale $u_3$.

To see explicitly how this setup relates the confinement scales $\Lambda_\textrm{\tiny{QCD}}$ and $\Lambda_D$, note that the fine-structure constants ($\alpha_i \equiv g_i^2/4\pi$) evaluated at the scale $u_3$ are related by
\begin{align}
    \frac{1}{\alpha_c} = \frac{1}{\alpha_1} &+ \frac{1}{\alpha_2} \,, \label{eq:Intro1} \\
    &\ \bigg\updownarrow \mathbb{Z}_2 \notag \\
    \frac{1}{\alpha_c} = \frac{1}{\alpha_1} &+ \frac{1}{\alpha_D} \,, \label{eq:Intro2}
\end{align}
where in the second line we have used the action of the exchange symmetry ($\alpha_2=\alpha_D$). Notice that $\alpha_c < \alpha_1,\, \alpha_D$ and as $\alpha_1$ is made larger, $\alpha_c$ approaches $\alpha_D$ from below. If the running of the ordinary and dark QCD couplings to lower energy is sufficiently similar, we then have $\Lambda_\textrm{\tiny{QCD}} < \Lambda_D$ which is the favoured hierarchy for dark QCD ADM. Furthermore, provided $\alpha_1(u_3)$ is not too small, we obtain the stronger and desired result $\Lambda_\textrm{\tiny{QCD}} \lesssim \Lambda_D$. As we shall discuss later, the $\mathbb{Z}_2$ must be either softly or spontaneously broken for phenomenological reasons; however, this does not qualitatively change the underlying mechanism described above.

The remainder of this paper is structured as follows. Section~\ref{sec:setup} describes the gauge symmetry structure and particle content of the model, and provides more details about how the confinement scales are related. The mass generation for the exotic fermions in the model, along with the symmetry breaking necessary for it, are discussed in Sec~\ref{sec:MassGeneration}. The scalar sector of the model and the spontaneous symmetry breaking dynamics are analysed in Sec.~\ref{sec:DSGBreaking}. In Sec.~\ref{sec:RunningCoupling}, the calculation of the running of the couplings, along with the resulting confinement scale ratios are presented. An axion solution to the strong CP problem is discussed in Sec.~\ref{sec:strongcp}. A brief discussion on possible baryogenesis mechanisms that may be used in completions of the model is presented in Sec.~\ref{sec:Baryogenesis}. We summarise our findings in Sec.~\ref{sec:Conclusion}, with further details provided in the Appendices.


\section{\label{sec:setup}The Model}

We consider a high energy theory that has an $SU(3)_1\times SU(3)_2 \times SU(2)_L \times U(1)_Y \times SU(3)_D$ local symmetry. Additionally, we impose a discrete $\mathbb{Z}_2$ symmetry that exchanges all fields (and the relevant couplings) charged under $SU(3)_2$ with their $SU(3)_D$ counterparts. The SM field content is extended by two new fermion fields, $\psi_2$ and $\psi_D$, which each have $N_f$ flavours, and by two bifundamental scalar fields $\Sigma_{12}$ and $\Sigma_{1D}$. The full field content for this model, along with the charge assignments under the gauge group, is given in Table \ref{tab:FieldContent}. The action of the $\mathbb{Z}_2$ exchange symmetry is 
\begin{equation}\label{eqn:Z2action}
    \psi_2 \leftrightarrow \psi_D \,,\quad \Sigma_{12}\leftrightarrow \Sigma_{1D} \,.
\end{equation}

\begin{table}[b]
\centering
\begin{tabular}{|c|c|c|c|c|c|c|c|}
    \hline
     Field & $SU(3)_1$ & $SU(3)_2$ & $SU(2)_L$ & $U(1)_Y$ & $SU(3)_D$ \\
     \hline
      $Q_L$ & 3 & 1 & 2 &\  \small{1/6} & 1\\ 
      $u_R$ & 3 & 1 & 1 &\  \small{2/3} & 1\\ 
      $d_R$ & 3 & 1 & 1 & -\small{1/3} & 1\\
      $l_L$ & 1 & 1 & 2 & -\small{1/2} & 1\\ 
      $e_R$ & 1 & 1 & 1 & -1 & 1\\ 
      H & 1 & 1 & 2 &\  \small{1/2} & 1\\
      \hline
      \rule{0pt}{2.5ex}$\Sigma_{12}$ & 3 & $\bar{3}$ & 1 &\  0 & 1\\
      $\Sigma_{1D}$ & 3 & 1 & 1 &\  0 & $\bar{3}$\\
      $\psi_{2}$ & 1 & 3 & 1 &\  0 & 1\\ 
      $\psi_{D}$ & 1 & 1 & 1 &\  0 & 3\\
      \hline
\end{tabular}
\caption{Field content and their quantum numbers under the extended gauge group.}
\label{tab:FieldContent}
\centering
\end{table}

The Lagrangian for the new field content is, 
\begin{equation}\label{eqn:FullSystemLagrangian}
\begin{split}    
    \mathcal{L}&\supset -\frac{1}{4}\sum_i(G^{a}_{i, \mu \nu})^2 + i\overline{\psi}_2 \slashed{D}\psi_2+ i\overline{\psi}_D \slashed{D}\psi_D\\
    &+\textrm{Tr}[(D_{\mu}\Sigma_{12})^{\dagger}D^{\mu}\Sigma_{12}]+ \textrm{Tr}[(D_{\mu}\Sigma_{1D})^{\dagger}D^{\mu}\Sigma_{1D}]\\
    &-\mathcal{L}_\psi-V \,,
\end{split}
\end{equation}
where $G_i^a$ are the different $SU(3)$ field-strength tensors, with $i=1,2,D$. Here, $\mathcal{L}_{\psi}$ describes the mass sector of the $\psi$ fermions, to be specified in Sec.~\ref{sec:MassGeneration}, and $V$ is the scalar potential.  

As previously mentioned, the spontaneous breaking $SU(3)_1 \times SU(3)_2 \to SU(3)_c$ leads to the relation in Eq.~\eqref{eq:Intro1} between the couplings at the breaking scale $u_3$. The action of the exchange symmetry then relates the QCD and dark QCD couplings. However, as will be argued in the next section, the $\mathbb{Z}_2$ symmetry must be broken. Hence, the precise relation depends on whether the exchange symmetry is broken above or below the scale $u_3$. If the exchange symmetry is unbroken at this scale, the relation between the visible and dark QCD couplings is established straightforwardly according to Eq.~\eqref{eq:Intro2}. However, if the exchange symmetry is broken at a scale higher than $u_3$, it is necessary to first evolve the couplings up to the scale where the exchange symmetry is restored, then use its action to relate the couplings.

The dark $SU(3)_c$ confines in the IR and the stable DM particles are bound states of the $\psi_D$ fields, in analogy to baryons in visible QCD. Guided by the ADM philosophy, we would like the ratio between the $SU(3)_c$ and $SU(3)_D$ confinement scales to be $\sim \mathcal{O}(1)$, as these scales determine the masses of the visible and dark baryons. 

The ratio of the confinement scales depends on both the number of flavours of new fermions, $N_f$, and the values of the gauge couplings in the UV. Given the field content, one can then solve the renormalisation group equations (RGEs), subject to the boundary condition in Eq.~\eqref{eq:Intro1}. It is convenient to use this condition to express the unknown couplings $\alpha_{1,2}(u_3)$ in terms of the QCD coupling $\alpha_c(u_3)$ according to
\begin{equation}\label{eqn:QCDpartitionxdefinition}
\begin{split}
    \alpha^{-1}_1(u_3)&=x \cdot\alpha_c^{-1}(u_3) \,, \\
    \alpha^{-1}_2(u_3)&=(1-x) \cdot\alpha_c^{-1}(u_3) \,,
\end{split}
\end{equation}
where $x$ is simply a number between 0 and 1, with the additional requirement that all couplings remain perturbative at the scale $u_3$ ($\alpha_i(u_3)<1$). This still leaves considerable freedom to choose the values of $\alpha_{1,2}$.

Notice that for small values of $x$, the $SU(3)_2$ and QCD couplings will roughly coincide at the scale $u_3$, while for $x\sim 1$ the couplings of $SU(3)_1$ and QCD approximately coincide. Given the exchange symmetry, which relates $\alpha_2$ and $\alpha_D$ in the UV, we favour the former case. However, while choosing different values for $x$ produces different results for the ratio of the confinement scales, we will show that the desired $\sim\mathcal{O}(1)$ ratio can be obtained for a wide range of $x$ values that are sufficiently smaller than one.


\section{\label{sec:MassGeneration} \texorpdfstring{$\mathbb{Z}_2$ Symmetry Breaking}{Z2 Symmetry Breaking}}

Within our framework, the DM mass is given by the confinement scale of dark QCD, $m_{\text{\tiny{DM}}}\sim \Lambda_D$, only if $\psi_D$ is massless or very light ($m_{\psi_D}\ll \Lambda_D$). Otherwise, the DM mass would be determined by the masses of the constituent fermions and no longer correlated with $\Lambda_\text{QCD}$. An important point arises from this requirement. The $\mathbb{Z}_2$ partner of $\psi_D$ is $\psi_2$, which is charged under QCD. Bounds from non-observation of coloured exotics at colliders~\cite{ATLAS:2022pib,CMS:2024nhn} imply that $\psi_2$ must be massive ($m_{\psi_2}\gtrsim 1.5$\,TeV). The exotic fermion mass sector must therefore break the $\mathbb{Z}_2$ symmetry in order to be consistent with observations and not spoil the overall framework. We propose two approaches to achieve the necessary $\mathbb{Z}_2$ breaking, as detailed in the following sections.

\subsection{\label{sec:MassGeneration/subsec:SoftSB}Soft Symmetry Breaking}

In principle, one could make both $\psi_2$ and $\psi_D$ massive, as long as $m_{\psi_D}\ll \Lambda_D$ and $m_{\psi_2}\gtrsim1.5$\,TeV. The simplest way to accomplish this is through the inclusion of soft $\mathbb{Z}_2$ symmetry breaking mass terms,
\begin{equation}\label{eqn:softSBMasses}
    \mathcal{L}_{\psi}=m_{\psi_D}\bar{\psi}_D\psi_D+m_{\psi_2}\bar{\psi}_2\psi_2 \,.
\end{equation}
The masses $m_{\psi_D}$ and $m_{\psi_2}$ are radiatively stable, free parameters of the theory and can be chosen to explicitly break the $\mathbb{Z}_2$ symmetry. Furthermore, because the breaking is soft, it does not disrupt the $\mathbb{Z}_2$ relations between the dimensionless couplings in the Lagrangian, most importantly between the gauge couplings $\alpha_2$ and $\alpha_D$.

While this is a viable approach, we focus our discussions around the more interesting case of spontaneous $\mathbb{Z}_2$ symmetry breaking in the rest of the text.

\subsection{\label{sec:MassGeneration/subsec:SSB}Spontaneous Symmetry Breaking}

The $\mathbb{Z}_2$-breaking mass terms for $\psi_{2}$ and $\psi_{D}$ can also be generated through spontaneous breaking of the $\mathbb{Z}_2$ by a gauge-singlet scalar field. However, if there were only one such scalar, the Yukawa couplings with the fermions would be identical in both sectors, making it impossible to generate $\mathbb{Z}_2$-breaking masses without also explicitly breaking the $\mathbb{Z}_2$ symmetry. Therefore, we must have more than one scalar involved.

The minimal viable option is to include two new complex scalar singlet fields, $\phi_2$ and $\phi_D$, and require that they only couple to the $\psi_2$ or $\psi_D$ sectors, respectively. In order to obtain $\mathbb{Z}_2$-breaking masses, we require that one of the scalar fields gets a vacuum expectation value (VEV) while the other one does not. 

This set-up implies the existence of two new global \textit{axial} symmetries, $U(1)_{2_A}$ and $U(1)_{D_A}$, as well as two \textit{vector} symmetries, $U(1)_{2_V}$ and $U(1)_{D_V}$, under which the $\psi$ and $\phi$ sectors are charged, as shown in Table \ref{tab:PQcharges}. Note that the axial symmetries forbid $\mathbb{Z}_2$-preserving mass terms for the fermions. The axial symmetries are, however, \textit{anomalous} under the corresponding $SU(3)$ gauge symmetries; this has important consequences for the strong-CP problem, as discussed in Sec.~\ref{sec:strongcp}. 

\begin{table}[t]
\centering
\begin{tabular}{|c|c|c|c|c|}
    \hline
     Field & $U(1)_{2_V}$& $U(1)_{2_A}$& $U(1)_{D_V}$ & $U(1)_{D_A}$ \\
     \hline
      $\psi_{2,L}$ &1&1&0&0\\
      $\psi_{2,R}$ &1&-1&0&0\\
      $\psi_{D,L}$ &0&0&1&1\\
      $\psi_{D,R}$ &0&0&1&-1\\
      $\phi_2$ &0&2&0&0\\
      $\phi_D$ &0&0&0&2\\
      \hline
\end{tabular}
\caption{Field content and their quantum numbers under the vector and axial $U(1)_2$ and $U(1)_D$ global symmetries. Note that the axial symmetries are anomalous. For the soft symmetry breaking approach in Eq.~(\ref{eqn:softSBMasses}), the axial symmetries are also softly broken.}
\label{tab:PQcharges}
\centering
\end{table}

The Yukawa Lagrangian for the $\psi$-fermion sector is thus
\begin{equation}\label{eqn:exoticYukawasector}
    \mathcal{L}_{\psi}=y_{\psi}\left(\phi_{D}\bar{\psi}_{D_L}\psi_{D_R}+\phi_2\bar{\psi}_{2_L}\psi_{2_R}\right)+\text{h.c.} \,.
\end{equation}
Hence, if $\phi_2$ gets a VEV and $\phi_D$ does not, we have a mechanism to generate the desired $\mathbb{Z}_2$-breaking masses in the fermion sector. 

\subsubsection{\texorpdfstring{$\phi$-Sector Spontaneous Symmetry Breaking}{phi-Sector Spontaneous Symmetry Breaking}}

Given the $\mathbb{Z}_2$ symmetry, the most general renormalisable potential for the $\phi$ sector is
\begin{equation}\label{eqn:phiPotential}
\begin{split}
    V_{\phi}=-m_{\phi}^2\left(\phi_D^*\phi_D +\phi_2^*\phi_2\right)&+\lambda_{\phi}\left[(\phi_D^*\phi_D)^2 +(\phi_2^*\phi_2)^2\right]\\
    &+\omega_{\phi}(\phi_D^*\phi_D\phi_2^*\phi_2) \,.
\end{split}
\end{equation}
This can be reparameterised in a useful way,
\begin{equation}
    V_{\phi}=\lambda_1\left(\phi_D^*\phi_D+\phi_2^*\phi_2-\frac{u_{\phi}^2}{2}\right)^2 +\lambda_2(\phi_D^*\phi_D\phi_2^*\phi_2) \,,
\end{equation}
where $\lambda_{1,2}$ and $u_{\phi}$ are real parameters. In the region where $\lambda_1,\lambda_2>0$, the global minimum is located where each term in brackets is individually zero, given that they are both positive definite. Therefore, the degenerate global minima are given by the VEV solutions
\begin{equation}\label{eqn:PQVEV1}
    \braket{\phi_D}=\frac{u_{\phi}}{\sqrt{2}}\,,\qquad \braket{\phi_2}=0 \,,
\end{equation}
and
\begin{equation}\label{eqn:PQVEV2}
    \braket{\phi_D}=0\,,\qquad \braket{\phi_2}=\frac{u_{\phi}}{\sqrt{2}} \,.
\end{equation}
These minima spontaneously break the $\mathbb{Z}_2$ and either $U(1)_{D_A}$ or $U(1)_{2_A}$. Since these minima are degenerate, the symmetry breaking produces stable domain walls. These present a cosmological problem that can be resolved simply by inflating the domain walls away (i.e.\ pre-inflationary $\mathbb{Z}_2$-breaking). After inflation, we identify our visible Universe as arising from a patch that fell into the minimum given by Eq.~(\ref{eqn:PQVEV2}).

Once $\phi_2$ acquires a VEV, we can parameterise it as
\begin{equation}\label{eqn:phipolarexpansion}
    \phi_2(x)=\frac{1}{\sqrt{2}}(u_{\phi}+\rho_2(x))e^{i \frac{a_2(x)}{u_{\phi}}} \,,
\end{equation}
where $\rho_2(x)$ is the radial degree of freedom, while $a_2(x)$ is the pseudo-Goldstone field of the $U(1)_{2_A}$ breaking. The latter can be identified with the QCD axion, as we discuss in Sec.~\ref{sec:strongcp}. The Yukawa Lagrangian then becomes
\begin{equation}\label{eqn:psimassSectorafterSSB}
    \mathcal{L}_{\psi}=\frac{y_{\psi}}{\sqrt{2}}(u_{\phi}+\rho_2)e^{i\frac{a_2}{u_{\phi}}}\bar{\psi}_{2_L}\psi_{2_R}+y_{\psi}\phi_{D}\bar{\psi}_{D_L}\psi_{D_R} \,.
\end{equation}
The mass of $\psi_2$ is
\begin{equation}\label{eqn:psi2mass}
    M_{\psi_2}=\frac{y_{\psi}}{\sqrt{2}}u_{\phi} \,,
\end{equation}
while $\psi_D$ remains massless. The unbroken vectorial $U(1)$ symmetries are identified as $\psi_2$- and $\psi_D$-number symmetries and ensure that bound states containing either $\psi_2$ or $\psi_D$ are stable.

Having a stable population of (coloured) $\psi_2$ fermions is problematic due to the stringent bounds on the abundance of fractionally charged hadrons~\cite{Perl:2009zz}. Therefore, when considering the cosmology of this model, it is necessary to either include a mechanism that makes $\psi_2$ unstable or have a reheating temperature lower than $M_{\psi_2}$.

The above symmetry breaking mechanism implies the existence of massless dark pions, which are the Goldstone bosons of the dark chiral symmetry. These can, however, be made massive by giving small masses to the dark fermions. In Appendix~\ref{app:DarkFermionMassGeneration} we discuss a possible completion of the scalar sector that generates small masses for the $\psi_D$ fermions while keeping the $\psi_2$ fermions heavy.

\vspace{1mm}
\section{\texorpdfstring{Symmetry Breaking In The Extended $SU(3)$ sector}{Symmetry Breaking In The Extended SU(3) sector}}
\label{sec:DSGBreaking}

The breaking of $SU(3)_1 \times SU(3)_2\rightarrow SU(3)_{1+2}\equiv SU(3)_{\text{\tiny{QCD}}}$  has been extensively studied in colouron models~\cite{Bai:2010dj,Bai:2017zhj,Chivukula:2013xka}. The field content of these models usually includes only one bifundamental scalar field, which is responsible for the symmetry breaking. When this bifundamental gets a VEV proportional to the identity, the unbroken symmetry corresponds to the diagonal subgroup of the original $SU(3)$ sector. In our model, due to the $\mathbb{Z}_2$ symmetry, the field content must include \textit{two} scalar bifundamental fields, $\Sigma_{12}$ and $\Sigma_{1D}$. Therefore, we must find a configuration of the bifundamental vacua corresponding to the desired breaking pattern $SU(3)_1\times SU(3)_2 \times SU(3)_D \rightarrow SU(3)_{\text{\tiny{QCD}}}\times SU(3)_D$.

The bifundamental fields transform under the $SU(3)$ symmetries according to
\begin{equation}\label{eqn:BFTransformations}
    \Sigma_{12} \to U_1 \Sigma_{12} U_2^\dagger\quad \textrm{and}\quad \Sigma_{1D} \to U_1 \Sigma_{1D} U_D^\dagger \,,
\end{equation}
where $U_i$ is a fundamental representation matrix for $SU(3)_i$ and each multiplet is represented as a $3 \times 3$ matrix. The most general scalar potential for the bifundamental fields is given by
\begin{widetext}
\begin{multline}\label{eqn:bifundamentalpotential}
     V_{\Sigma}= -m_{\Sigma}^2\left(\text{Tr}(\Sigma_{12} \Sigma_{12}^{\dagger})+\text{Tr}(\Sigma_{1D} \Sigma_{1D}^{\dagger})\right)+\frac{\lambda_{\Sigma}}{2}\left( [\text{Tr}(\Sigma_{12} \Sigma_{12}^{\dagger})]^2+[\text{Tr}(\Sigma_{1D} \Sigma_{1D}^{\dagger})]^2\right)\\
     +\frac{\kappa_{\Sigma}}{2}\left(\text{Tr}(\Sigma_{12} \Sigma_{12}^{\dagger}\Sigma_{12} \Sigma_{12}^{\dagger}) +\text{Tr}(\Sigma_{1D} \Sigma_{1D}^{\dagger}\Sigma_{1D} \Sigma_{1D}^{\dagger})\right)
     - \mu_{\Sigma}\,\big(\text{det}(\Sigma_{12})+\text{det}(\Sigma_{1D})+\text{h.c.}\big)\\
     +\omega_{\Sigma}\,\text{Tr}(\Sigma_{12}\Sigma_{12}^{\dagger})\text{Tr}(\Sigma_{1D}\Sigma_{1D}^{\dagger})+\sigma_{\Sigma}\,\text{Tr}(\Sigma_{12}^{\dagger}\Sigma_{1D}\Sigma_{1D}^{\dagger}\Sigma_{12}) +V_{\phi\Sigma}+V_{H\Sigma}\,,
\end{multline}
\end{widetext}
where $V_{\phi\Sigma}$ and $V_{H\Sigma}$ contain the mixing terms with the $\phi_i$ and Higgs sectors, respectively. We note that important tree-level corrections coming from $V_{\phi\Sigma}$ modify the mass terms differently for each bifundamental, as shown in Eq.~(\ref{eqn:BFmassSplitting}). (The cubic couplings receive small modifications at higher order, an effect we neglect.)

In order to achieve the desired breaking, this potential must have a global minimum such that $\Sigma_{12}$ acquires a VEV proportional to the identity matrix, while $\Sigma_{1D}$ has a zero VEV.

The VEVs of the bifundamentals are general $3\times3$ complex matrices which can be expressed as,
\begin{equation}
    \braket{\Sigma_{1i}}=A_{i} D_i B_i,\quad \text{for }i=2,D,
\end{equation}
where $A_i$ and $B_i$ are general unitary matrices, and $D_i$ is diagonal. Using the three different $SU(3)$ symmetries to redefine the bifundamental fields, it is possible to remove three of the four unitary matrices appearing in the VEVs, while introducing an overall phase. Therefore, without loss of generality, we can write the VEVs as
\begin{align}\label{eqn:2BFGeneralVEV}
    \braket{\Sigma_{12}}=
    &\begin{bmatrix}
        s_1&0&0\\
        0&s_2&0\\
        0&0&s_3
    \end{bmatrix} e^{i\alpha/3} \,,\\
    \braket{\Sigma_{1D}}=U\cdot
    &\begin{bmatrix}
        s_4&0&0\\
        0&s_5&0\\
        0&0&s_6
    \end{bmatrix} e^{i\beta/3} \,,
\end{align}
where $s_i\geq0$, $\alpha,\beta\in\{-\pi,\pi\}$, and $U$ is a general unitary matrix.

The $\mathbb{Z}_2$ symmetry breaking (either soft or spontaneous) allows for the masses of the bifundamentals to be different. We then find that there is indeed a significant region of parameter space where the global minimum corresponds to the desired symmetry breaking pattern with VEVs
\begin{equation}\label{eqn:BFZ2breakingVEV}
    \braket{\Sigma_{12}}=\frac{u_3}{\sqrt{6}}\cdot \mathbb{I}\,,\qquad \braket{\Sigma_{1D}}=0 \,,
\end{equation}
where
\begin{equation}\label{eqn:u3definition}
    u_3=\frac{\sqrt{3}}{\sqrt{2}(3\lambda_{\Sigma}+\kappa_{\Sigma})}\left(\mu_{\Sigma}\pm\sqrt{\mu_{\Sigma}^2+4m_{12}^2(3\lambda_{\Sigma}+\kappa_{\Sigma})}\right) \,,
\end{equation}
and $m_{12}$ is the mass of $\Sigma_{12}$. We show explicitly how this is achieved and discuss the resulting particle spectrum in Appendix~\ref{app:BFSymmetryBreaking}.

\phantom{x}

\section{\label{sec:RunningCoupling}Running of the Couplings and Confinement Scale Ratios}

\begin{figure*}[!t]
    \includegraphics[width=0.4\textwidth]{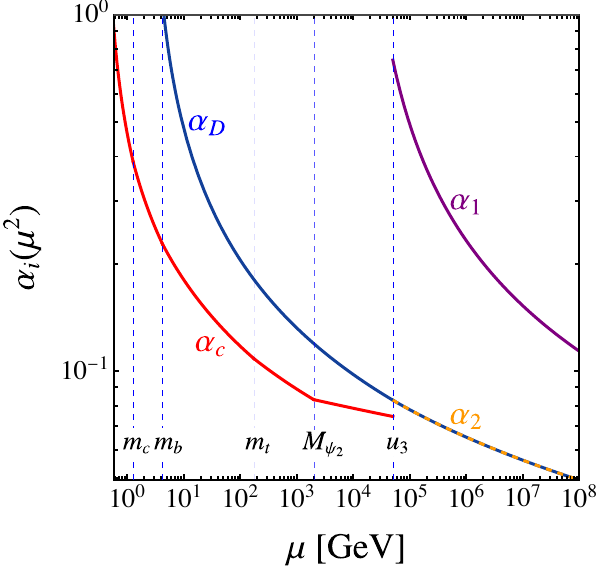}
    \hspace{1cm}
    \includegraphics[width=0.4\textwidth]{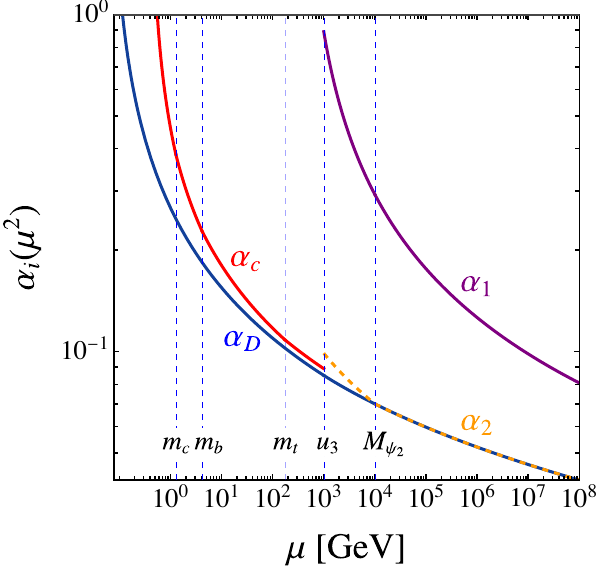}
    \caption{Running of the coupling constants for benchmark points in the two configurations, $M_{\psi_2}<u_3$ (left panel) and $M_{\psi_2}>u_3$ (right panel). The running of $\alpha_c$ (red), $\alpha_1$ (purple), $\alpha_2$ (dashed yellow) and $\alpha_D$ (dark blue) are shown up to a high scale, taken as $10^8$ GeV. For the left panel, the benchmark values are $u_3=10^{4.7}$ GeV and $M_{\psi_2}=10^{3.3}$ GeV, while for the right panel they are taken as $u_3=10^{3}$ GeV and $M_{\psi_2}=10^4$ GeV. These parameter choices were made to showcase the behaviour of the couplings for the different configurations.}
    \label{fig:RunningSixFields}
\end{figure*}

As explained in Sec.~\ref{sec:setup}, the coupling constants of visible and dark QCD are related. We now show how this relation predicts the dark confinement scale, thus providing an explanation for the similarity of the masses of the bound states of visible and dark QCD across a significant region of parameter space.

The RGEs for the different $SU(3)$ couplings are coupled due to the presence of the bifundamental fields $\Sigma$. At two-loop, the $\beta$-functions also depend on the quark and $\psi$ Yukawa couplings.\footnote{There are no $\psi$ Yukawa couplings for the softly-broken $\mathbb{Z}_2$ case.} However, as will be argued in Sec.~\ref{sec:strongcp/subsec:axionpheno}, the coupling $y_\psi$ must be small and hence its contribution to the running is negligible. We also neglect all quark Yukawa couplings except for the top quark. The two-loop $\beta_1(\alpha_1,\alpha_2,\alpha_D,y_t)$ function for $SU(3)_1$ is then~\cite{Jones:1981we,Machacek:1983fi}
\begin{widetext}
\begin{equation}\label{eqn:betafunc}
\begin{split}
    \beta_1(\{\alpha_i\},y_t)&= \frac{\alpha_1^2}{4\pi}\left( \frac{4}{3} T(\boldsymbol{R_f})n_f+\frac{1}{3}T(\boldsymbol{R_s})n_s -\frac{11}{3}C_2(G)\right)\\
    &+\frac{\alpha^3_1}{16\pi^2}\left(2\left(\frac{10}{3}C_2(G)+2C_2(\boldsymbol{R_f})\right)T(\boldsymbol{R_f})n_f+\left(\frac{2}{3}C_2(G))+4C_2(\boldsymbol{R_s})\right)T(\boldsymbol{R_s})n_s-\frac{34}{3}C_2^2(G)\right)\\
    &+\frac{\alpha_1^2 \alpha_2}{16 \pi^2}\left(4C_2(\boldsymbol{R_s})T(\boldsymbol{R_s})n_s\right)+\frac{\alpha_1^2 \alpha_D}{16 \pi^2}\left(4C_2(\boldsymbol{R_s})T(\boldsymbol{R_s})n_s\right)-\frac{\alpha_1^2 y_t}{16\pi^2}\left(4\ T(\boldsymbol{R_f})\right),
\end{split}
\end{equation}
\end{widetext}
where $n_f$ and $n_s$ are the number of \textit{active} fermions and complex scalars charged under $SU(3)_1$, respectively. Here, $T(\boldsymbol{R})$ is the Dynkin index, which for the fundamental representation is $T(\boldsymbol{N})=1/2$, $C_2(G)=N$ is the quadratic Casimir for the adjoint representation, and $C_2(\boldsymbol{R_s})=C_2(\boldsymbol{R_f})=\frac{N^2-1}{2N}$ are the quadratic Casimirs for the scalars and fermions in the fundamental representation. To obtain the $SU(3)_2$ or $SU(3)_D$ $\beta$-function one simply exchanges $1\leftrightarrow 2$ or $1\leftrightarrow D$, respectively, and then remove terms proportional to the product $\alpha_2 \alpha_D$ or $y_t$.

We solve the coupled two-loop RGEs for $\alpha_1,\alpha_2$ and $\alpha_D$ together with the one-loop RGE for $y_t$. Particle mass thresholds are accounted for via the one-loop matching conditions
\begin{equation}
    \alpha_i^{N'}(\mu_{M})=\left( 1-\frac{\alpha_i(\mu)}{6\pi}\left(n'_f+\frac{1}{4}n'_s\right)\ln\left(\frac{\mu}{M}\right) \right) \alpha_i^{N}(\mu_{M}) \,,
\end{equation}
where $\mu_{M}$ is the matching scale, $M$ is the mass of the decoupling states, and $\alpha^{N}$, $\alpha^{N'}$ are the couplings with $N$ and $N'$ active fields, respectively. $n'_f$ and $n'_s$ are the number of fermion or scalar fields decoupling, respectively, and thus $N'=N-(n_f'+n_s')$. For simplicity, we fix $\mu_{M}=M$ but note that our results will vary slightly depending on the choice of the matching scale.

Lastly, we need to connect the values of the running couplings in the IR with the masses of the visible and dark sector baryons. These are approximately of order the respective confinement scales, $\Lambda_i$, which we take to be the scales at which $\alpha_i(\Lambda_i)=1$.

\subsection{\label{subsec:ScaleRatios}Ratio of Confinement Scales}

\begin{figure*}[!t]
    \includegraphics[width=0.45\textwidth]{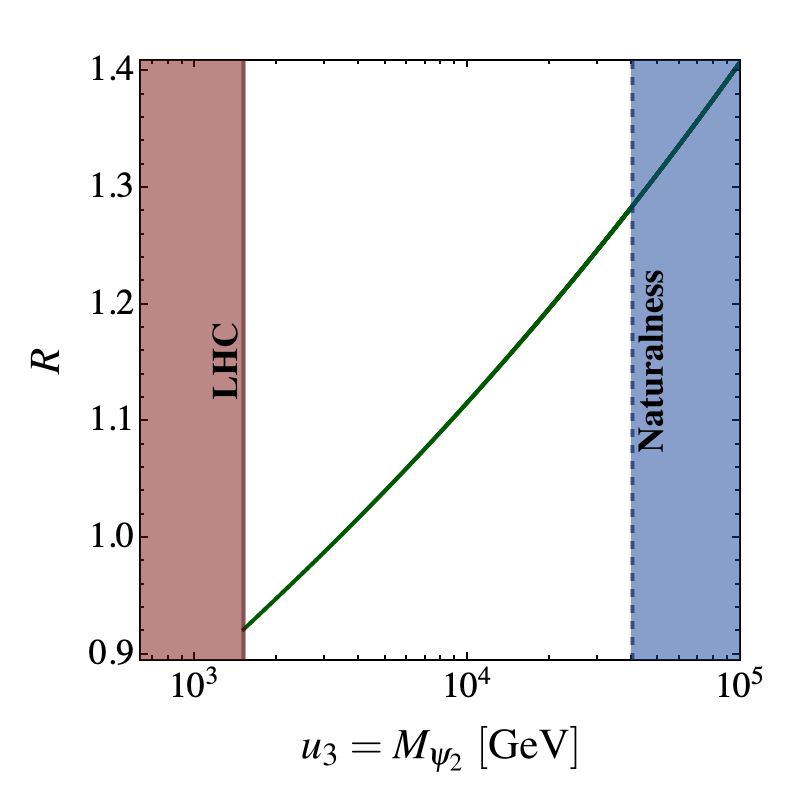}
    \hfill
    \includegraphics[width=0.45\textwidth]{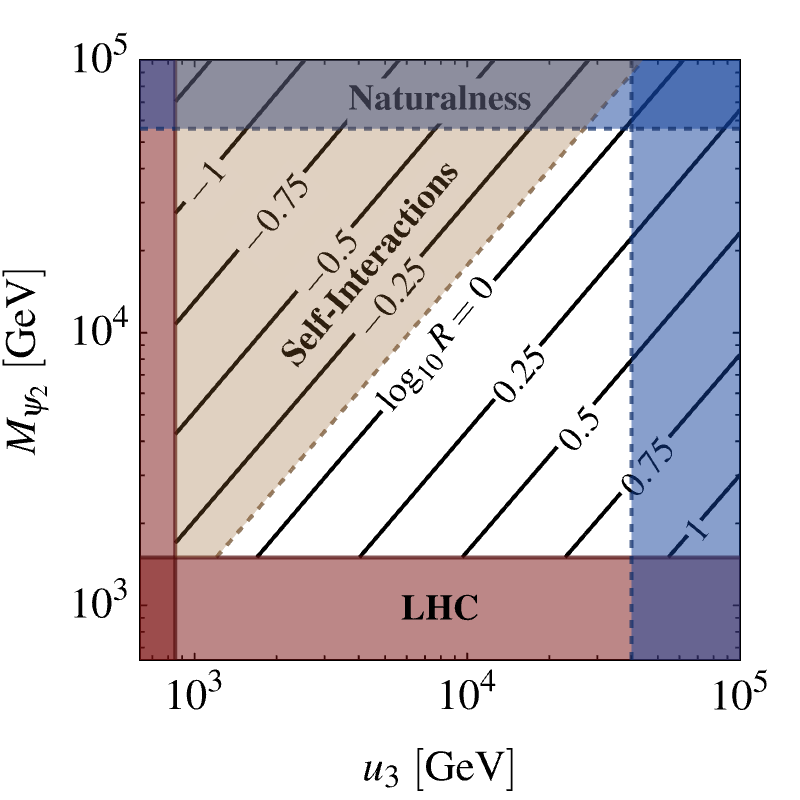}
    
    \caption{Ratio of the visible and dark confinement scales, $R=\Lambda_D/\Lambda_{\text{\tiny{QCD}}}$. The left panel shows $R$ as a function of $u_3$ for the special case where $u_3=M_{\psi_2}$. The right panel shows the results as contours of $R$ in the ($u_3,M_{\psi_2})$ plane. Regions where $R \lesssim 0.8$ are excluded by astrophysical bounds on self-interacting DM and are shown as brown shaded regions. The blue shaded regions are disfavoured by electroweak naturalness and the red shaded regions are excluded by collider experiments.}
    \label{fig:ScaleRatiosSixDarkFields}
\end{figure*}

The calculation of the confinement scales depends on the number of $\psi$ fermions, $N_f$, and the values of the gauge couplings in the UV, which are determined by the parameter $x$ in Eq.~(\ref{eqn:QCDpartitionxdefinition}). For the results presented here, we choose $N_f=6$, as this mirrors the number of quarks in the SM, and $x=0.1$, which imposes that $\alpha_2$ and $\alpha_c$ roughly coincide at $u_3$. We note that these choices are arbitrary and different values will lead to different results; we discuss these possibilities at the end of this section.

The evolution of the couplings, and hence the values of the confinement scales, also depend on the free parameters $M_{\psi_2}$ and $u_3$. There are two distinct scenarios, depending on whether the $\psi_2$ fermions decouple above or below the $SU(3)$ breaking scale, $M_{\psi_2} \geq u_3$ or $M_{\psi_2} \leq u_3$. The running of the gauge couplings is qualitatively different in the two cases, as shown in Fig.~\ref{fig:RunningSixFields}. First, note that in both cases the running of $\alpha_2$ and $\alpha_D$ is identical at energies $\mu>\max\{ M_{\psi_2},u_3\}$ and hence the action of the exchange symmetry can be used to relate the couplings at these scales.

For the configuration $M_{\psi_2}\leq u_3$, shown in the left panel of Fig.~\ref{fig:RunningSixFields}, we can use Eq.~(\ref{eq:Intro2}) at the symmetry breaking scale $u_3$ to determine the value of $\alpha_D$ in the UV. We then evolve it down to calculate the dark confinement scale. For energies $M_{\psi_2}<\mu<u_3$, the visible sector has \textit{double} the number of active fermions compared to the dark sector (recall that $\psi_2$ is a fundamental of $SU(3)_c$), and hence evolves at a slower rate. After crossing the $M_{\psi_2}$ threshold, the visible and dark sectors have the same number of active flavours until the SM quarks start to decouple in the visible sector, causing it to evolve faster.

Conversely, for the configuration $M_{\psi_2}\geq u_3$, shown in the right panel of Fig.~\ref{fig:RunningSixFields}, we cannot use Eq.~(\ref{eq:Intro2}) at the scale $u_3$ to determine $\alpha_D$. Instead, we must first evolve $\alpha_2$ up to $M_{\psi_2}$ where we can then relate it to $\alpha_D$ using the exchange symmetry. Then, we calculate the evolution of the dark sector to obtain the confinement scale. In this scenario, for the region $u_3<\mu<M_{\psi_2}$, $SU(3)_2$ has only the scalar bifundamental fields contributing to its running, while $SU(3)_D$ has both the bifundamental fields and the fermions $\psi_D$. As such, $SU(3)_2$ runs significantly faster than $SU(3)_D$. However, once the $u_3$ threshold is crossed, both the visible and dark QCD sectors have the same running behaviour, up to effects from SM quark thresholds.

We show the results for the ratio between the confinement scales, $R=\Lambda_D/\Lambda_{\text{\tiny{QCD}}}$, for both $M_{\psi_2}\leq u_3$ and $M_{\psi_2}\geq u_3$ in the right panel of Fig.~\ref{fig:ScaleRatiosSixDarkFields}. Notice that in the region where $M_{\psi_2}\leq u_3$, larger separations between $M_{\psi_2}$ and $u_3$ lead to greater ratios, as this extends the region $M_{\psi_2}\leq \mu \leq u_3$ where QCD runs slower than dark QCD. On the contrary, as the separation gets smaller, the ratio decreases and approaches $\sim 1$ when $M_{\psi_2}\sim u_3$. For configurations where $M_{\psi_2}\geq u_3$ it is observed that for increasing separation between the scales, the resulting ratio decreases, contrary to the previous case. This can be explained by the fact that, in the region $u_3\leq\mu\leq M_{\psi_2}$, the running of $\alpha_2$ is much faster than that of $\alpha_D$. Therefore, by extending this region, $\alpha_2$, which effectively sets the value of $\alpha_c$ using Eq.~(\ref{eqn:QCDpartitionxdefinition}), can become large, leading to smaller ratios. Conversely, as the separation between scales decreases, the ratio increases, again approaching $R\sim1$ for regions where $M_{\psi_2}\sim u_3$. Lastly, the left panel of Fig.~\ref{fig:ScaleRatiosSixDarkFields} shows the special case where $u_3=M_{\psi_2}$. This further illustrates that when the two scales are (approximately) equal, the ratio between the scales is $R\gtrsim 1$ and shows little variation with $u_3$. 

In the regions of parameter space where $\Lambda_D < \Lambda_{QCD}$, the DM self-interaction cross section can exceed the neutron self-interaction cross section. Such large cross sections are constrained by astrophysical bounds on DM self-interactions, with observations of colliding galaxy clusters excluding $\sigma_\text{DM}/m_\text{DM} < 0.5\,\text{cm}^2/\text{g}$~\cite{Harvey:2015hha}. The DM mass and self-interaction cross section are approximately given by $m_\text{DM}\sim m_n R$ and $\sigma_\text{DM}\sim\sigma_{n}/R^2$, respectively, with $\sigma_{n}\simeq10^{-24}\,\text{cm}^2$ the neutron cross section and $m_n$ the neutron mass\footnote{This scaling assumes that the dark pions obtain masses of order $\Lambda_D$, as discussed in Appendix~\ref{app:DarkFermionMassGeneration}. If the $\psi_D$ are strictly massless, then the dark nucleons will have long-range interactions and thus invalidate this scaling approximation.}. Under these assumptions, self-interacting DM bounds exclude the region of parameter space where $R\lesssim0.8$. This region is shaded in brown in Fig.~\ref{fig:ScaleRatiosSixDarkFields}.

Collider searches provide lower bounds on both the mass of $\psi_2$ and the scale $u_3$, since the latter sets the masses of the heavy coloured scalars and vectors in the model. First, note that $\psi_2$ will hadronise, forming a massive, stable hadron. The CMS search for stable SUSY R-hadrons (containing either a gluino or a squark) can therefore be reinterpreted to constrain this model. The bounds on gluino R-hadrons exclude masses up to $2.06$ TeV, while top squark R-hadron searches exclude masses up to $1.5$ TeV~\cite{CMS:2024nhn}. The bound on the mass of our colour-triplet fermion $\psi_2$ is expected to lie between these. As a conservative bound, we take $M_{\psi_2}\gtrsim 1.5$\,TeV. Similarly, searches for heavy gluons constrain the $SU(3)$ symmetry breaking scale. The breaking leads to a field spectrum containing heavy gluons, called colourons, and colour-octet and colour-singlet scalars. Further details are provided in Appendix~\ref{app:BFSymmetryBreaking}. The exact limits on $u_3$ depend on the mass spectrum of this sector. It was shown in Refs.~\cite{Bai:2018jsr,Dobrescu:2007yp} that for the case where the colouron is heavier than the scalar octet and singlet, collider searches result in the bound $u_3 \gtrsim0.85$\,TeV. However, different mass spectra will lead to different bounds. For example, if the colouron is lighter than the scalars then dijet searches will likely lead to stronger bounds on this scale~\cite{CMS:2019gwf}. The regions excluded by collider searches are shown in red in Fig.~\ref{fig:ScaleRatiosSixDarkFields}.

Naturalness of the electroweak scale can be used to impose upper bounds on the masses of the new fermions and scalars in the model. The upper bounds on the masses of new coloured fermions or scalars that interact with the SM purely via $SU(3)_c$ gauge interactions are $\mathcal{O}(10-100)$ TeV~\cite{Clarke:2016jzm}. In this work we take the naturalness bounds $M_{\psi_2}<56$\,TeV and $u_3<40$\,TeV.\footnote{In terms of the notation in Ref.~\cite{Clarke:2016jzm} these bounds correspond to the fine tuning measure satisfying $\Delta<100$.} These are shown in blue in Fig.~\ref{fig:ScaleRatiosSixDarkFields}. However, it is important to note that while collider constraints represent strict bounds on the parameters, naturalness driven constraints are subjective: the limit depends on what is deemed a tolerable degree of fine-tuning.

We find that the model contains a significant region of parameter space where the ratio between the confinement scales is an $\mathcal{O}(1-10)$ number and is greater than unity. Notably, in the regions also favoured by naturalness, $R\sim\mathcal{O}(1)$, making it a very attractive solution. Additionally, this region is also the most accessible to future collider searches. 

As stated before, the choice of both $N_f$, and the value of $x$ are arbitrary. We explore different choices for these parameters in the Appendices. The results for the minimal flavour set-up, $N_f=1$, are shown in Appendix~\ref{app:MWE}. Results for different choices of $x$ are shown in Appendix~\ref{app:xPlots}. Note that we restrict $x>0.1$ to ensure that $\alpha_1$ remains perturbative for all values of $u_3$ we consider. Significantly, the result is that values of $x$ in the range $0.1 - 0.4$ can produce a dark confinement scale that is less than a factor of ten higher than the QCD scale and so are deemed acceptable for explaining the cosmological coincidence. We have effectively reduced an unquantifiable but very surprising coincidence to one that occurs naturally in a significant region of the parameter space. That constitutes success.


\section{Strong CP Problem and the Axion}\label{sec:strongcp}

Upon closer inspection, the mass generation mechanism described in Sec.~\ref{sec:MassGeneration/subsec:SSB} has very important consequences for the strong CP problem. Initially, the extended $SU(3)_1 \times SU(3)_2 \times SU(3)_D$ gauge group means that the Lagrangian contains three topological $\theta$ terms,
\begin{multline}\label{eqn:topologicalSector}
    \mathcal{L} \supset \theta_1 \frac{g_1^2}{32\pi^2}\tilde{G}_{1}^{a, \mu \nu}G_{1\, a, \mu \nu}+\theta_2 \frac{g_2^2}{32\pi^2}\tilde{G}_{2}^{a, \mu \nu}G_{2\, a, \mu \nu}\\
    +\theta_D \frac{g_D^2}{32\pi^2} \tilde{G}_{D}^{a, \mu \nu}G_{D\, a, \mu \nu} \,,
\end{multline}
where $G_i^{\mu\nu}$ are the field strength tensors of the different $SU(3)$ groups and $\tilde{G}^{\mu\nu}=\frac{1}{2}\epsilon^{\mu \nu \alpha \beta}G_{\alpha\beta}$. The $\mathbb{Z}_2$ symmetry enforces $\theta_2=\theta_D$ and $g_2=g_D$.

The $U(1)_{2_A}$ and $U(1)_{D_A}$ global symmetries are anomalous under $SU(3)_2$ and $SU(3)_D$, respectively. Since the $\psi_D$ fermions are massless, the $\theta_D$ term is unphysical and can be ``rotated away'' by performing a $U(1)_{D_A}$ transformation.\footnote{For massive $\psi_D$ fields, the $U(1)_{D_A}$ symmetry is broken and thus $\theta_D$ can not be ``rotated away''. However, if the masses are generated spontaneously, the $U(1)_{D_A}$ symmetry is identified as a dark-PQ symmetry and there is a dark axion that dynamically sets $\theta_D$ to zero.} In the visible sector, on the other hand, the $U(1)_{2_A}$ symmetry is spontaneously broken and the $\psi_2$ fermions become massive.

It is convenient to perform an $a_2$-dependent field-redefinition of $\psi_2$ to remove the coupling with $a_2$ in Eq.~(\ref{eqn:psimassSectorafterSSB}). This introduces a coupling to $\tilde{G}_2 G_2$ via the path integral Jacobian, with the topological sector Lagrangian becoming
\begin{multline}\label{eqn:topologicalSectoraxion}
    \mathcal{L}\supset\frac{N_f g_2^2}{32\pi^2}\frac{(a_2(x)+\theta_2)}{u_{\phi}}\tilde{G}_2^{a, \mu\nu} G_{2\, a, \mu\nu}\\
    +\theta_1\frac{g_1^2}{32\pi^2}\tilde{G}_1^{a, \mu\nu} G_{1\, a, \mu\nu} \,.
\end{multline}
Note that in this basis the field $a_2(x)$ also has derivative, axial-vector couplings with the $\psi_2$ fields.
As the Universe cools, $SU(3)_1\times SU(3)_2$ breaks down to $SU(3)_c$, with the QCD gluons, $G_c$, and the colourons, $G_H$, given by~\cite{Bai:2010dj,Bai:2018jsr},
\begin{equation}\label{eqn:GluonDefinition}
\begin{split}
    G_c^{\mu}&=\cos \delta \ G_1^{\mu}+\sin \delta\ G_2^{\mu} \,,\\
    G_H^{\mu}&=\sin \delta \ G_1^{\mu}-\cos \delta\ G_2^{\mu} \,,
\end{split}
\end{equation}
where $\tan \delta=g_1/g_2$. Thus, the topological sector can be expressed at low-energies as, 
\begin{equation}\label{eqn:axionGGcoupling}
    \mathcal{L}\supset \frac{N_f g_c^2}{32\pi^2}\frac{(a_2(x)+\theta_c)}{u_{\phi}}\tilde{G}_c^{a,\mu\nu} G_{c\, a,\mu\nu} \,,
\end{equation}
where we have integrated out the heavy fields $G_H$. The $\theta$ term associated with QCD is thus,
\begin{equation}
    \theta_c=\theta_1+\theta_2 \,,
\end{equation}
and the $a_2(x)$ field is identified as the QCD axion.

In this sense, the $U(1)_{2_A}$ symmetry can be identified as the Peccei-Quinn (PQ) symmetry~\cite{Weinberg:1977ma,Wilczek:1977pj,Peccei:1977hh,Peccei:1977ur} in axion solutions of the strong CP problem. More precisely, the relevant field content corresponds to that of a Kim-Shifman-Vainshtein-Zakharov (KSVZ) axion model~\cite{Kim:1979if, Shifman:1979if}. Thus, the inclusion of the mass generation mechanism in Sec.~\ref{sec:MassGeneration/subsec:SSB} provides an axion solution to the strong CP problem.

In this model, the axion also couples to the massive colourons, which have been integrated out in Eq.~\eqref{eqn:axionGGcoupling}. One might be concerned that this results in an additional (non-QCD) contribution to the axion potential that could lead to an axion quality problem. However, the only explicit breaking of the $U(1)_{2_A}$ PQ-symmetry is due to the chiral anomaly; the axion potential therefore arises purely via non-perturbative effects through the coupling to the $SU(3)_2$ topological term in Eq.~\eqref{eqn:topologicalSectoraxion}. Above the $SU(3)_1 \times SU(3)_2$ breaking scale, $u_3$, there is a short distance $SU(3)_2$ instanton contribution to the axion potential. The size of this contribution is $V \sim y_\psi u_3^4 \exp{(-8\pi^2/g_2^2)}$, where $g_2$ is evaluated at the scale $u_3$. Since $SU(3)_2$ is asymptotically free, this contribution can easily be suppressed relative to the QCD contribution to the potential, $V_{QCD} \sim m_\pi^2 f_\pi^2$. 

\subsection{Axion Phenomenology}\label{sec:strongcp/subsec:axionpheno}

It is clear from Eq.~(\ref{eqn:axionGGcoupling}) that $u_{\phi}/N_f$ is identified as the axion decay constant, $f_a$. There are stringent lower bounds on the value of $f_a$, in particular, derived from astrophysical observations~\cite{Dolan:2022kul,Buschmann:2021juv,Capozzi:2020cbu,Springmann:2024mjp}. The strongest of these is from neutron star cooling~\cite{Buschmann:2021juv} and implies 
\begin{equation}\label{eqn:uphilowerbound}
    u_{\phi}\gtrsim 10^8\,\text{GeV}.
\end{equation}
This bound\footnote{Of course, if the $\mathbb{Z}_2$ is softly broken as in Sec.~\ref{sec:MassGeneration/subsec:SoftSB}, there is no axion and this bound does not apply.} is many orders of magnitude stronger than those set by collider searches for the exotics that were discussed in Sec.~\ref{subsec:ScaleRatios}. 

The above bound also has implications for the mass of the fermion $\psi_2$, leading to
\begin{equation}\label{eqn:psi2lowmassapprox}
    M_{\psi_2}=\frac{y_{\psi}}{\sqrt{2}}u_{\phi} \gtrsim \frac{y_{\psi}}{\sqrt{2}}\ 10^8 \text{ GeV}.
\end{equation}
As discussed in Sec.~\ref{subsec:ScaleRatios}, naturalness of the electroweak scale implies $M_{\psi_2} \lesssim 40\,$TeV, which requires $y_{\psi} \lesssim \mathcal{O}(10^{-4})$. Note that such a small Yukawa coupling is, however, technically natural in the sense of 't Hooft~\cite{tHooft:1979rat}, since there is an additional $U(1)$ symmetry in the limit $y_{\psi} \to 0$.

The axion is also a viable DM candidate; however, in the present model we do not want the axion to be the DM. We thus need a cosmological history which \textit{under-produces} the axion relic abundance. Due to the arguments in Sec.~\ref{sec:MassGeneration/subsec:SSB}, the PQ-breaking scale $u_{\phi}$ should be greater than the reheating temperature and we have a pre-inflationary axion scenario. The axion abundance is then produced by the misalignment mechanism~\cite{Preskill:1982cy,Abbott:1982af,Dine:1982ah} leading to $\Omega_a h^2\approx 0.12 \left(\frac{\theta_i}{2.155}\right)^2 \left(\frac{f_a}{2\times 10^{11} \text{ GeV}}\right)^{1.16}$, with $\theta_i$ the initial misalignment angle~\cite{OHare:2024nmr,Borsanyi:2016ksw}. Lower values of $f_a=u_\phi/N_f$ are therefore favoured to ensure that the axion constitutes only a small fraction of the DM.


\section{\label{sec:Baryogenesis}Baryogenesis Mechanisms}

A complete explanation of the relation in Eq.~(\ref{eq:cosmocoincidence}) must provide both a connection between the masses of visible and dark particles and a connection between their number densities. In this work, we have introduced a new mechanism that can explain the former but have not yet touched on the latter. While it would be interesting to construct a complete model that can generate the necessary asymmetries in the visible and dark sectors, this goes beyond the scope of the present work. Here, we briefly discuss how the model could be extended to introduce a connection between asymmetries in the two sectors.

In the context of ADM, a number density relating/reprocessing operator is an operator which violates both baryon (or lepton) and dark baryon number, but preserves a linear combination of them. One such portal is the lepton-dark baryon portal,
\begin{equation}\label{eqn:LeptonPortal}
    \frac{1}{M_N^3}\bar{l}_L H(\epsilon^{a'b'c'}\overline{\psi}_{D_L, a'}\psi_{D_R, b'}\psi_{D_R, c'}) \,,
\end{equation}
where $M_N$ is the scale of the effective interaction, and $a',b',c'$ are dark-colour indices.\footnote{It is also possible to construct a neutron-dark-baryon portal, where the conserved symmetries are $U(1)_{3B-D_V}$ and $U(1)_{3B+D_A/3}$.} We immediately see that the combinations $U(1)_{L+D_V}$ and $U(1)_{L-D_A/3}$ are conserved, where $L$ represents global lepton number in the SM and $D_V$ and $D_A$ represent the global vector and axial symmetries introduced in Sec.~\ref{sec:MassGeneration/subsec:SSB}.

This portal will reprocess lepton number into dark baryon number. Naturally, the connection to thermal leptogenesis can be made~\cite{Fukugita:1986hr}. However, at this point we leave a detailed analysis of this possibility for future work.


\section{Conclusion}\label{sec:Conclusion}

In this work, we have constructed a new asymmetric dark matter mechanism that relates the confinement scale of a dark QCD sector with the visible one. The dark matter candidate is a baryon-like particle composed of dark fermion fields. In analogy to QCD, the mass of this state is given by the dark confinement scale. Therefore, this mechanism provides an explanation for why the dark matter mass is related to the proton mass. 

The model establishes a relation between the dark and visible sector gauge couplings through spontaneous symmetry breaking in an extended $SU(3)$ sector, combined with a $\mathbb{Z}_2$ exchange symmetry. This relation implies that the visible coupling is comparable to, but lower than, the dark one at a particular UV scale. Evolving these couplings to low energies, we then obtain a relation between their respective confinement scales. 

Our results show a considerable parameter region where the ratios between the confinement scales are $\mathcal{O}(1)$, implying that the mass of the DM particle is similar to that of the proton, $m_\text{DM}\sim m_p$. The full region of viable parameter space exhibits a range of possible values for the ratio, $\mathcal{O}(1-10)$, making it an adaptable approach for relating $m_\text{DM}$ and $m_p$. The mechanism is also testable, with the new particle content potentially accessible at future collider experiments through searches for the massive gluons and other exotic coloured particles. Furthermore, embedded in the model is a KSVZ-type axion solution to the strong CP problem, solving a second SM shortcoming.  

We have shown that this extension of the colour sector provides an interesting solution to the relation between the masses of visible and asymmetric dark matter. It would be interesting to explore how this model can be incorporated into a full cosmological treatment of the coincidence problem, including a well-defined baryogenesis mechanism and evolution of the relic densities.

\begin{acknowledgments}
We thank Yi Chung for helpful comments on an earlier version of this paper. This work was supported in part by the Australian Research Council through the ARC Centre of Excellence for Dark Matter Particle Physics CE200100008. P.C. is supported by the Australian Research Council Discovery Early Career Researcher Award DE210100446.
\end{acknowledgments}

\bibliography{references}

\appendix

\section{Generation of the dark fermion mass}\label{app:DarkFermionMassGeneration}

In this appendix, we describe two possible mechanisms to generate a small mass for the dark fermions $\psi_D$, making the dark pions massive. 

The simplest approach is to add a soft $\mathbb{Z}_2$-breaking mass term for the $\psi_D$ fields. In this case, the mass can be chosen to be as small as desired and is radiatively stable. 

However, we find that a minimal extension to the mechanism presented in Sec.~\ref{sec:MassGeneration/subsec:SSB} is able to generate the $\psi_D$ mass spontaneously. We discuss this now.

In addition to the scalar sector in Sec.~\ref{sec:MassGeneration/subsec:SSB}, we consider an additional complex scalar field, $\varphi$, which only carries charge under the global ($U(1)_{2_A}$, $U(1)_{D_A}$) symmetries, $\varphi \sim (-2,-2)$, and is its own $\mathbb{Z}_2$ partner. 

The potential for this extended scalar sector is,
\begin{equation}
\begin{split}
    V=&-m_{\phi}^2 (\phi_2^* \phi_2+\phi_D^* \phi_D)+\lambda_{\phi}((\phi_2^* \phi_2)^2+(\phi_D^* \phi_D)^2)\\
    &+\omega_{\phi}(\phi_2^* \phi_2)(\phi_D^* \phi_D)-m_{\varphi}^2 \varphi^*\varphi+\lambda_{\varphi}(\varphi^*\varphi)^2\\
    &+\epsilon (\varphi\phi_2 \phi_D+\varphi^*\phi_2^* \phi_D^*)
    +\omega_{\varphi}(\phi_2^* \phi_2+\phi_D^* \phi_D)\varphi^*\varphi \,.
\end{split}
\end{equation}
In order to see how the mechanism works, let us treat the trilinear term as a small perturbation to the potential. Such a small $\epsilon$ is technically natural, as there is an additional $U(1)$ symmetry in the limit $\epsilon\rightarrow 0$. In the $\epsilon=0$ limit, the potential can be factorised as per
\begin{equation}
\begin{split}
    V = \alpha& \left(\phi_2^*\phi_2+\phi_D^*\phi_D +\varphi^*\varphi -\frac{u_{\phi}^2}{2} -\frac{u_{\varphi}^2}{2}\right)^2\\
    +&\, \beta\left(\varphi^*\varphi-\frac{u_{\varphi}^2}{2}\right)^2 + \kappa\left(\phi_2^*\phi_2+\phi_D^*\phi_D-\frac{u_{\phi}^2}{2}\right)^2\\
    +&\, \lambda \phi_2^*\phi_2\phi_D^*\phi_D \,.
\end{split}
\end{equation}
In the region where $\alpha,\beta,\kappa,\lambda>0$, the VEV structure is either
\begin{equation} \label{eq:phi2_VEV}
    \braket{\phi_2}=\frac{u_{\phi}}{\sqrt{2}}\,,\quad \braket{\phi_D}=0\,,\quad \braket{\varphi}=\frac{u_{\varphi}}{\sqrt{2}} \,,
\end{equation}
or
\begin{equation}
    \braket{\phi_2}=0\,,\quad \braket{\phi_D}=\frac{u_{\phi}}{\sqrt{2}}\,,\quad \braket{\varphi}=\frac{u_{\varphi}}{\sqrt{2}} \,,
\end{equation}
where
\begin{align}
    u_\phi &= \sqrt{\frac{\lambda_{\varphi}m_{\phi}^2-\frac{1}{2}\omega_{\varphi}m_{\varphi}^2}{\lambda_{\phi}\lambda_{\varphi}-\frac{1}{4}\omega^2_{\varphi}}} \,,\\
    u_{\varphi} &= \sqrt{\frac{\lambda_{\phi}m_{\varphi}^2-\frac{1}{2}\omega_{\varphi}m_{\phi}^2}{\lambda_{\phi}\lambda_{\varphi}-\frac{1}{4}\omega^2_{\varphi}}} \,.
\end{align}
As in the main text, we consider the first configuration \eqref{eq:phi2_VEV}, which leads to massive exotic fermions ($m_{\psi_2} > 0$).

Now, we consider the corrections to these VEVs due to the trilinear term, which induces a non-zero VEV for $\phi_D$. In the limit $\epsilon \ll m_\phi, m_\varphi$ this is approximately given by
\begin{equation}
    \braket{\phi_D}\simeq -\epsilon\frac{ u_{\varphi}u_{\phi}}{-m_{\phi}^2+ \omega_{\phi}u_{\phi}^2+\omega_{\varphi}u_{\varphi}^2} \,.
\end{equation}
Notice that in this limit the VEVs satisfy the hierarchy $|\langle\phi_D\rangle|\ll\braket{\phi_2},\braket{\varphi}$. For this VEV configuration, the Yukawa sector in Eq.~(\ref{eqn:exoticYukawasector}) implies that both $\psi_2$ and $\psi_D$ obtain masses. Furthermore, we can make $\braket{\phi_2}$ sufficiently large to ensure $m_{\psi_2}$ obeys the observational constraints on exotic coloured particles, while simultaneously keeping $\braket{\phi_D}$ small,  ensuring $m_{\psi_D}$ is smaller than the dark confinement scale.

\section{\texorpdfstring{$SU(3)$-sector symmetry breaking}{SU(3) sector symmetry breaking}}\label{app:BFSymmetryBreaking}

In this appendix, we show that there exists parameter space which results in the desired symmetry breaking pattern $SU(3)_1 \times SU(3)_2 \to SU(3)_c$, with the bifundamental VEVs given by Eq.~\eqref{eqn:BFZ2breakingVEV}. First, it is useful to rewrite the scalar potential of the $SU(3)$ sector in the form
\begin{multline} \label{eq:SigmaPot}
     V_{\Sigma}= V_0(\Sigma_{12}) + V_0(\Sigma_{1D}) + \omega_{\Sigma}\, \text{Tr}(\Sigma_{12}\Sigma_{12}^{\dagger})\text{Tr}(\Sigma_{1D}\Sigma_{1D}^{\dagger}) \\
     +\sigma_{\Sigma}\,\text{Tr}(\Sigma_{12}^{\dagger}\Sigma_{1D}\Sigma_{1D}^{\dagger}\Sigma_{12}) +V_{\phi\Sigma}+V_{H\Sigma}\,,
\end{multline}
where $V_0$ includes terms that depend on only a single bifundamental field,
\begin{multline}
    V_0(\Sigma) = -m_{\Sigma}^2\, \text{Tr}(\Sigma \Sigma^{\dagger}) + \frac{\lambda_{\Sigma}}{2}\, [\text{Tr}(\Sigma \Sigma^{\dagger})]^2 \\ 
    + \frac{\kappa_{\Sigma}}{2}\, \text{Tr}(\Sigma \Sigma^{\dagger} \Sigma \Sigma^{\dagger}) - \mu_{\Sigma}\, (\text{det}(\Sigma)+\text{h.c.}) \,.
\end{multline}
In the following, we neglect $V_{H\Sigma}$ since its effects on the $\Sigma$ vacuum can be easily suppressed and do not affect the overall discussion.

As discussed in the main text, the VEVs can be written, without loss of generality, as
\begin{align}\label{eqn:2BFGeneralVEV_app}
    \braket{\Sigma_{12}}=
    &\begin{bmatrix}
        s_1&0&0\\
        0&s_2&0\\
        0&0&s_3
    \end{bmatrix} e^{i\alpha/3} \,,\\
    \braket{\Sigma_{1D}}=U\cdot
    &\begin{bmatrix}
        s_4&0&0\\
        0&s_5&0\\
        0&0&s_6
    \end{bmatrix} e^{i\beta/3} \,. \label{eqn:2BFGeneralVEV_app_1D}
\end{align}
Substituting these VEVs into the potential, the $V_0$ terms become 
\begin{multline}\label{eqn:singleBFpotential}
    V_0(\braket{\Sigma_{12}})=-m_{\Sigma}^2(s_1^2+s_2^2+s_3^2)+\frac{\lambda_{\Sigma}}{2}(s_1^2+s_2^2+s_3^2)^2\\
    +\frac{\kappa_{\Sigma}}{2}(s_1^4+s_2^4+s_3^4)-2\mu_{\Sigma}s_1s_2s_3 \cos\alpha \,,
\end{multline}
and similarly for $\braket{\Sigma_{1D}}$. 

If the contributions to the potential from $V_{\phi\Sigma}$ are neglected, the $\mathbb{Z}_2$ symmetry in the $\Sigma$ sector is unbroken. The parameters entering the two $V_0$ potentials ($m_\Sigma$, $\lambda_\Sigma$, $\kappa_\Sigma$, $\mu_\sigma$) are therefore the same. Hence, in the region of parameter space where $\omega_\Sigma=\sigma_\Sigma=0$, the global minimum of the full potential is where the diagonal matrices in Eq.~(\ref{eqn:2BFGeneralVEV_app}) are in the same configuration. However, including the effects of $V_{\phi \Sigma}$ and the mixing terms (non-zero $\omega_\Sigma$ and/or $\sigma_\Sigma$) can change this conclusion. In particular, tree-level corrections from
\begin{multline}\label{eqn:phiSigmaPotential}
    V_{\phi\Sigma}=\lambda_{\phi\Sigma}\left(\phi_2^{\dagger}\phi_2 \text{Tr}(\Sigma^{\dagger}_{12}\Sigma_{12})+\phi_D^{\dagger}\phi_D \text{Tr}(\Sigma^{\dagger}_{1D}\Sigma_{1D})\right) \\
    +\lambda_{\phi\Sigma}' \left(\phi_D^{\dagger}\phi_D \text{Tr}(\Sigma^{\dagger}_{12}\Sigma_{12})+\phi_2^{\dagger}\phi_2 \text{Tr}(\Sigma^{\dagger}_{1D}\Sigma_{1D})\right) \,.
\end{multline}
generate different masses for each bifundamental once $\phi_2$ obtains a VEV. We discuss the details of the mass splitting in the next section and, for the moment, focus only on the effect this has on the $\Sigma$ VEV structure.

We focus on the case where the $\mathbb{Z}_2$ symmetry is broken \textit{before} $SU(3)_1\times SU(3)_2$. We can then integrate out $\phi_2$ and $\phi_D$ to obtain an effective potential for $\Sigma_{12}$ and $\Sigma_{1D}$, which can be written as
\begin{multline} \label{eq:SigmaPotEff}
     V_{\Sigma}^{\text{eff}}= V^{12}_0(\Sigma_{12}) + V^{1D}_0(\Sigma_{1D}) + \omega_{\Sigma}\, \text{Tr}(\Sigma_{12}\Sigma_{12}^{\dagger})\text{Tr}(\Sigma_{1D}\Sigma_{1D}^{\dagger}) \\
     +\sigma_{\Sigma}\,\text{Tr}(\Sigma_{12}^{\dagger}\Sigma_{1D}\Sigma_{1D}^{\dagger}\Sigma_{12}) \,,
\end{multline}
where the $V_0^{1i}$ still depend on only a single bifundamental field, but the mass terms have now been modified by the corrections coming from $V_{\phi\Sigma}$,
\begin{multline}
    V_0^{1i}(\Sigma)=-m_{1i}^2\, \text{Tr}(\Sigma \Sigma^{\dagger}) + \frac{\lambda_{\Sigma}}{2}\, [\text{Tr}(\Sigma \Sigma^{\dagger})]^2 \\ 
    + \frac{\kappa_{\Sigma}}{2}\, \text{Tr}(\Sigma \Sigma^{\dagger} \Sigma \Sigma^{\dagger}) - \mu_{\Sigma}\, (\text{det}(\Sigma)+\text{h.c.}) \,.
\end{multline}
As before, focusing on the region where $\omega_{\Sigma}=\sigma_{\Sigma}=0$, the global minimum is given by configurations that separately minimise $V_0^{12}(\braket{\Sigma_{12}})$ and $V_0^{1D}(\braket{\Sigma_{1D}})$. However, it is now possible for $\braket{\Sigma_{12}}$ and $\braket{\Sigma_{1D}}$ to be in different configurations, depending on the parameter region. In particular, from the single bifundamental vacuum analysis in Ref~\cite{Bai:2017zhj}, we know that the parameter region where the global minimum of $V^{1D}_0(\braket{\Sigma_{1D}})$ is given by the trivial VEV is 
\begin{equation}\label{eqn:1DparamRegion}
    m_{1D}^2<0 \text{ and} -\mu^2_{\Sigma}/m^2_{1D}<4(3\lambda_{\Sigma}+\kappa_{\Sigma}) \,.
\end{equation}
Conversely, in the regions,
\begin{equation}\label{eqn:12paramRegion1}
 m^2_{12}>0 \text{ and } \{\kappa_{\Sigma}\geq0 \text{ or } \frac{\mu_{\Sigma}}{m_{12}}>\frac{-\kappa_{\Sigma}}{\sqrt{\lambda_{\Sigma}+\kappa_{\Sigma}}}\} \,,
\end{equation}
or
\begin{equation}\label{eqn:12paramRegion2}
    m_{12}^2<0 \text{ and } \frac{9}{2}(3\lambda_{\Sigma} +\kappa_{\Sigma})<-\mu^2_{\Sigma}/m^2_{12} \,,
\end{equation}
the global minimum of $V^{12}_0(\braket{\Sigma_{12}})$ is given by $\braket{\Sigma_{12}}=u_3\cdot \mathbb{I}_{3\times3}$, where
\begin{equation}\label{eqn:u3definition_app}
    u_3=\frac{\sqrt{3}}{\sqrt{2}(3\lambda_{\Sigma}+\kappa_{\Sigma})}\left(\mu_{\Sigma}\pm\sqrt{\mu_{\Sigma}^2+4m_{12}^2(3\lambda_{\Sigma}+\kappa_{\Sigma})}\right) \,.
\end{equation}

So far, including only the $V_0$ sector of the potential, we have shown that having a different mass term for each bifundamental allows us to find a parameter region where the minimum of the potential corresponds to the desired symmetry breaking pattern. However, we must now consider the effect of the mixing terms. Substituting the general VEVs into the effective potential, it can be factorised into the form 
\begin{multline}\label{eqn:2BFVEVPot}               
    V_\Sigma^{\text{eff}}=V^{12}_0(\braket{\Sigma_{12}})+V^{1D}_0(\braket{\Sigma_{1D}})\\
    +\omega_{\Sigma} (s_1^2+s_2^2+s_3^2)(s_4^2+s_5^2+s_6^2)\\
    +\sigma_{\Sigma} \Big(s_1^2(\Delta_{1}^2)
    +s_2^2(\Delta_2^2)+s_3^2(\Delta_3^2)\Big) \,,
\end{multline}
where $\Delta_i^2=s_4^2|a_{i1}|^2+s_5^2|a_{i2}|^2+s_6^2|a_{i3}|^2$, with $a_{ij}$ the elements of the unitary matrix $U$ in Eq.~\eqref{eqn:2BFGeneralVEV_app_1D}. Notice that if the parameter space is restricted to $\omega_{\Sigma},\sigma_{\Sigma}>0$, both mixing terms are positive definite. Furthermore, these terms vanish when $\braket{\Sigma_{1D}}=0$ (i.e. $s_{4,5,6}=0$). The desired symmetry breaking configuration with $\braket{\Sigma_{1D}}=0$ and $\braket{\Sigma_{12}}=u_3\cdot \mathbb{I}_{3\times3}$ is therefore a global minimum  of the potential in the region where $\omega_{\Sigma},\sigma_{\Sigma}>0$ is satisfied together with Eq.~(\ref{eqn:1DparamRegion}) and either~(\ref{eqn:12paramRegion1}) or~(\ref{eqn:12paramRegion2}).

\subsection{\texorpdfstring{Mass corrections from $\mathbb{Z}_2$ breaking}{Mass corrections from Z2 breaking}}\label{app:BFSymmetryBreaking/subsec:MassCorrectionGen}

Here, we show explicitly how the bifundamental mass splitting is achieved. Consider again $V_{\phi\Sigma}$\,, shown in Eq.~(\ref{eqn:phiSigmaPotential})\,. Once $\phi_2$ gets a VEV, it provides corrections to the mass terms for both the $\Sigma_{12}$ and $\Sigma_{1D}$ fields.\footnote{Naturally, if the $\mathbb{Z}_2$ symmetry was instead softly broken, we can simply add the desired mass corrections to the potential. In this case there is no $V_{\phi\Sigma}$ potential, but the analysis presented in the previous section otherwise holds.} However, the mass terms depend on different coupling constants, $\lambda_{\phi\Sigma}$ and $\lambda_{\phi\Sigma}'$. As such, the effective mass terms of the bifundamentals are
\begin{equation}\label{eqn:BFmassSplitting}
\begin{split}
    m_{12}^2&=m^2_{\Sigma}+\frac{\lambda_{\phi\Sigma} u_{\phi}^2}{2} \,,\\
    m_{1D}^2&=m^2_{\Sigma}+\frac{\lambda_{\phi\Sigma}' u_{\phi}^2}{2} \,.
\end{split}
\end{equation}
The $\mathbb{Z}_2$-breaking contributions are proportional to $u_{\phi}$, which has a lower bound of $10^8$ GeV, as explained in Sec.~\ref{sec:strongcp/subsec:axionpheno}. Therefore, unless $\lambda_{\phi\Sigma}$ and $\lambda_{\phi\sigma}'$ are small, these drive the masses to large values; from Eq.~(\ref{eqn:u3definition_app}), this also means $u_3\sim m_{12}\sim u_{\phi}$. However, the limit $\lambda_{\phi\Sigma},\lambda_{\phi\Sigma}',y_{\psi}\rightarrow0$ is a technically natural one. We can therefore consider the parameter space in which $\lambda_{\phi\Sigma},\lambda_{\phi\sigma}' \ll 1$, where the masses of both bifundamentals remain of order $m_\Sigma$ and we can have $u_3 \ll u_\phi$.

Combining the above expressions for the effective bifundamental masses with the conditions for the desired symmetry breaking in Eqs.~(\ref{eqn:1DparamRegion}),~(\ref{eqn:12paramRegion1}) and~(\ref{eqn:12paramRegion2}) yields the constraints
\begin{equation}\label{eqn:Z21DBFVEVcond}
    \frac{\lambda_{\phi\Sigma}' u_{\phi}^2}{2} <-m_{\Sigma}^2, \text{ and } \frac{\lambda_{\phi\Sigma}' u_{\phi}^2}{2} <-\frac{\mu^2_{\Sigma}}{4(3\lambda_{\Sigma}+\kappa_{\Sigma})}-m^2_{\Sigma} \,,
\end{equation}
and for $m_{12}^2$,
\begin{equation}
    \frac{\lambda_{\phi\Sigma}u_{\phi}^2}{2}<-m_{\Sigma}^2, \text{ and } \frac{\lambda_{\phi\Sigma}u_{\phi}^2}{2}>-\frac{\mu_{\Sigma}^2}{\frac{9}{2}(3\lambda_{\Sigma}+\kappa_{\Sigma})}-m^2_{\Sigma} \,,
\end{equation}
or
\begin{equation}
    \frac{\lambda_{\phi\Sigma}u_{\phi}^2}{2}>-m_{\Sigma}^2, \text{ and } \kappa_{\Sigma}>0 \,.
\end{equation}
These conditions can be easily satisfied across a large fraction of the parameter space. While the combination of electroweak naturalness and axion constraints favours the parameter space where $u_3\ll u_{\phi}$, we note that the desired $\mathbb{Z}_2$ breaking VEV configuration can also be achieved when $u_3 \sim u_\phi$.

\subsection{Asymptotic Conditions}

To ensure that the scalar potential is bounded from below, we must impose additional conditions on the parameters. The full scalar potential is given by Eqs.~(\ref{eqn:bifundamentalpotential}) and~(\ref{eqn:phiPotential}), with only the quartic terms relevant in the large field limit. For this potential, a straightforward argument can be made; by restricting the parameter region to $\kappa_{\Sigma},\lambda_{\Sigma},\omega_{\Sigma},\sigma_{\Sigma},\lambda_{\phi},\omega_{\phi}>0$, it is immediately clear that the potential is always bounded from below. This choice can be made consistently with the conditions needed for the desired VEV configuration discussed in the previous section and in section~\ref{sec:MassGeneration/subsec:SSB}.

\subsection{Field Spectrum}

The $SU(3)$ breaking described above leads to a rich new spectrum of fields. This has been widely studied in colouron models~\cite{Bai:2010dj} and we only list the relevant results here.

Once $\Sigma_{12}$ acquires a VEV as in Eq.~(\ref{eqn:BFZ2breakingVEV}), we can expand around the vacuum as
\begin{equation}
    \Sigma_{12}=\frac{u_3+\phi_R+i\phi_I}{\sqrt{6}}\mathbb{I}+(\Theta_H^a+i\Theta_G^a)T^a \,.
\end{equation}
Here, $T^a$ are the $SU(3)_c$ generators, $\Theta_G^a$ are the Nambu-Goldstone bosons that get eaten by the gauge bosons, $\Theta_H^a$ is a massive colour octet and $\phi_{I,R}$ are colour singlets. The gauge bosons that become massive are commonly referred to as colourons. The masses of these new states are
\begin{equation}
\begin{split}
     M_{\phi_I}^2&=\sqrt{\frac{3}{2}}\mu_{\Sigma}u_3 \,,\\
    M_{\phi_R}^2&=\frac{1}{3}\left((\kappa_{\Sigma}+3\lambda_{\Sigma})u_3^2 -M_{\phi_I}^2)\right) \,,\\
    M_{\Theta_H}^2&=\frac{1}{3}\left(2M_{\phi_I}^2+\kappa_{\Sigma}u_3^2\right) \,,\\
    M_{\text{\tiny{colouron}}}&=\sqrt{\frac{2}{3}}\frac{g_c}{\sin 2\delta}u_3 \,,
\end{split}
\end{equation}
where $\tan\delta=g_1/g_2$.

Of the above states, those potentially relevant for the running of the QCD coupling are $\Theta_H^a$ and the colorouns, which can safely be taken to have a mass $M_{\Theta_H},M_{\text{\tiny{colouron}}}\sim u_3$, assuming $\mu_{\Sigma},\kappa_{\Sigma}\sim \mathcal{O}(1)$. Thus, these states do not contribute to the evolution of $\alpha_c$ at lower energies.

In addition to this spectrum, we also have the complex bifundamental $\Sigma_{1D}$ field, which does not acquire a VEV, but contributes to the running of both $SU(3)_1$ and $SU(3)_D$. We assume that $m_{1D}\sim m_{12}$ and hence $m_{1D}$ is roughly $\mathcal{O}(u_3)$. As such, $\Sigma_{1D}$ does not contribute to the running of the gauge groups below $u_3$.

\section{\texorpdfstring{Minimal Model ($N_{f}=1$)}{Minimal Model (Nf=1)}}\label{app:MWE}

In this appendix, we present results for the ratio between the visible and dark confinement scales for the case where the field content only includes \textit{one} flavour of $\psi_2$ and $\psi_D$ fields, $N_f=1$. 

\begin{figure*}[t]
    \includegraphics[width=0.4\textwidth]{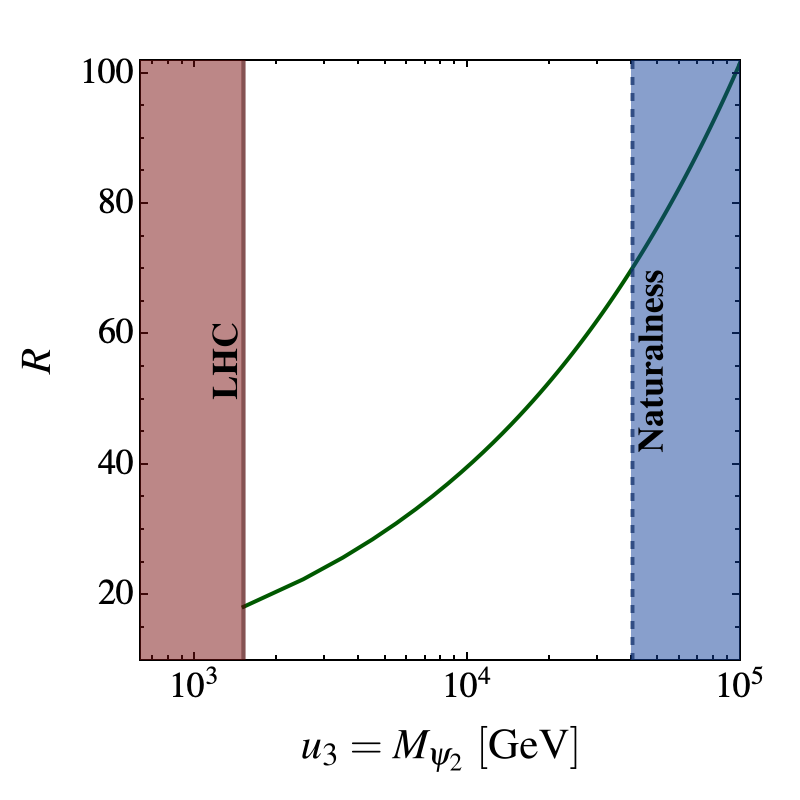}
    \hfill
    \includegraphics[width=0.4\textwidth]{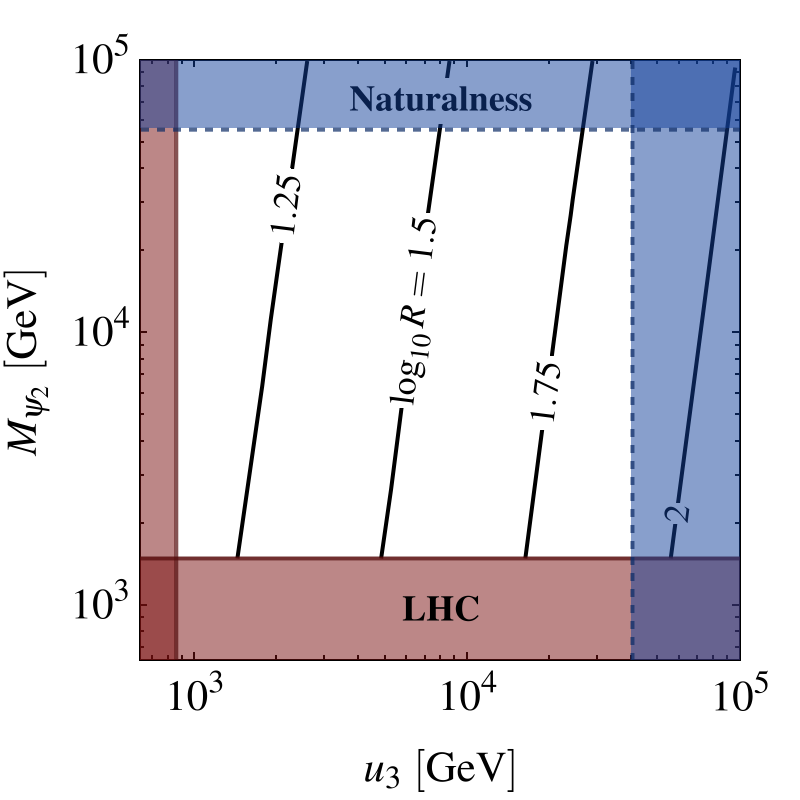}
    
    \caption{Ratio of the visible and dark confinement scales, $R=\Lambda_D/\Lambda_{\text{\tiny{QCD}}}$, for $N_f=1$. The left panel shows $R$ as a function of $u_3$ for the special case where $u_3=M_{\psi_2}$. The right panel shows the results as contours of $R$ in the ($u_3,M_{\psi_2})$ plane. The blue and red shaded regions correspond to the naturalness and collider bounds, respectively, as discussed in the main text.}
    \label{fig:ScaleRatiosMWE}
\end{figure*}

The main difference between this case and  the one presented in the main text is the fact that dark QCD has only one fermion contributing to its running, while QCD still has the six SM flavours, as well as the $\psi_2$ field (however this only contributes at scales $\mu>M_{\psi_2}$). Therefore, as soon as the $SU(3)$ sector breaks down to QCD, the dark sector runs significantly faster than the visible one. Here, the threshold at $M_{\psi_2}$ only decreases the number of active flavours for QCD by one, so its impact is not as significant as in the $N_f=6$ case. Hence, this means that the ratio of confinement scales primarily depends on the $SU(3)$ breaking scale, $u_3$, and only slightly varies with respect to $M_{\psi_2}$. The results are shown in Fig.~\ref{fig:ScaleRatiosMWE}. Across the entire allowed parameter space, the dark sector confines at a significantly higher scale than visible QCD, resulting in values of $R$ that are significantly greater than those obtained in the $N_f=6$ case.

\section{\texorpdfstring{Ratio of confinement scales for different $x$}{Ratio of scales for different x}}\label{app:xPlots}

\begin{figure*}[t]
    \includegraphics[width=0.3\textwidth]{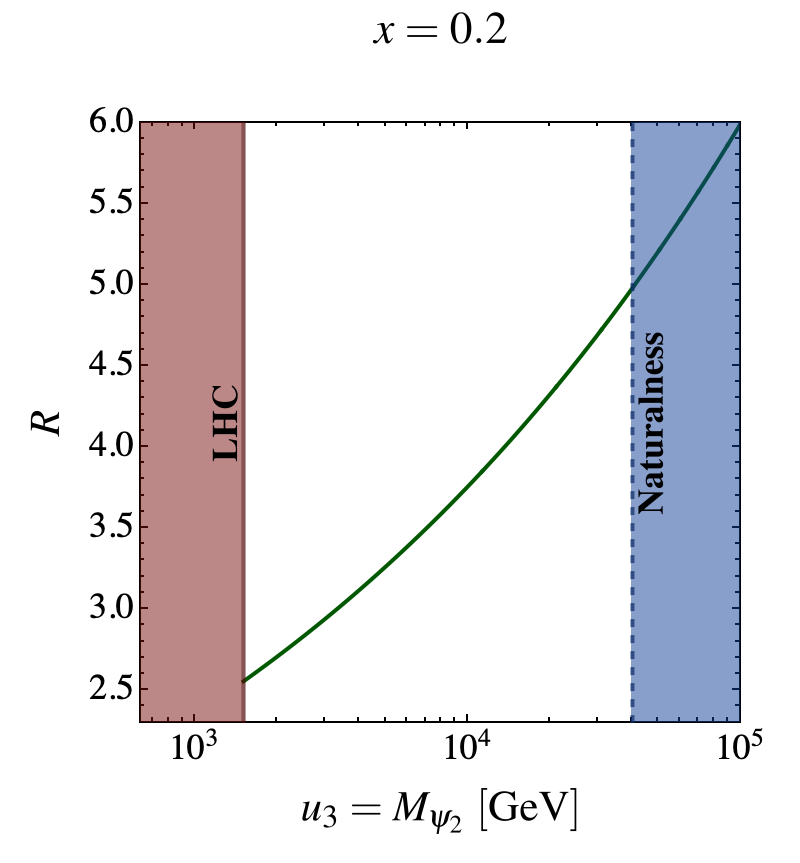}
    \hfill
    \includegraphics[width=0.3\textwidth]{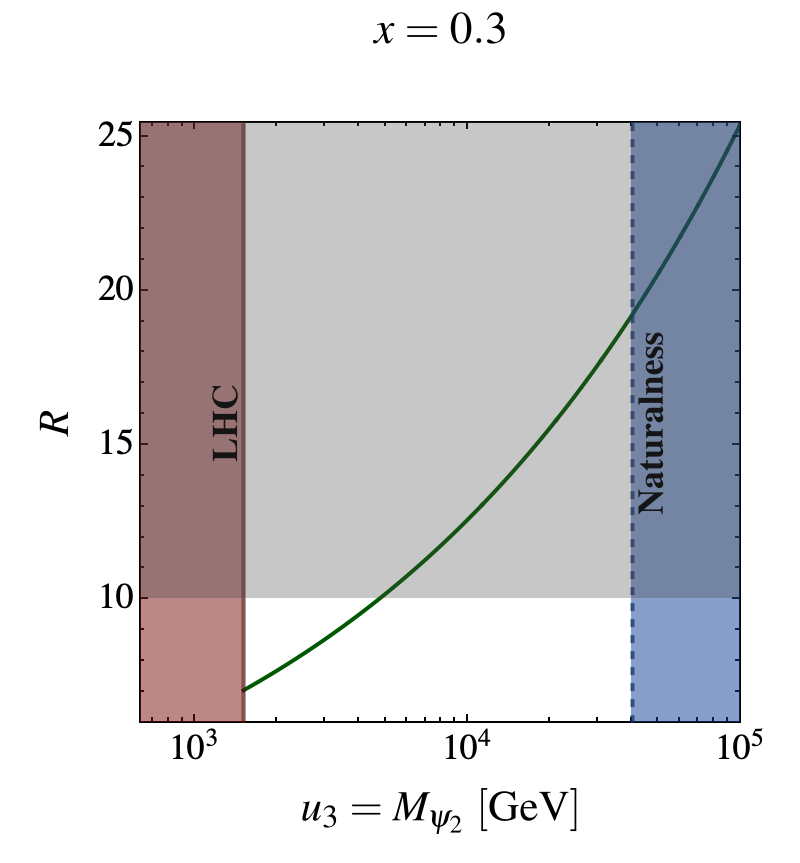}
    \hfill
    \includegraphics[width=0.3\textwidth]{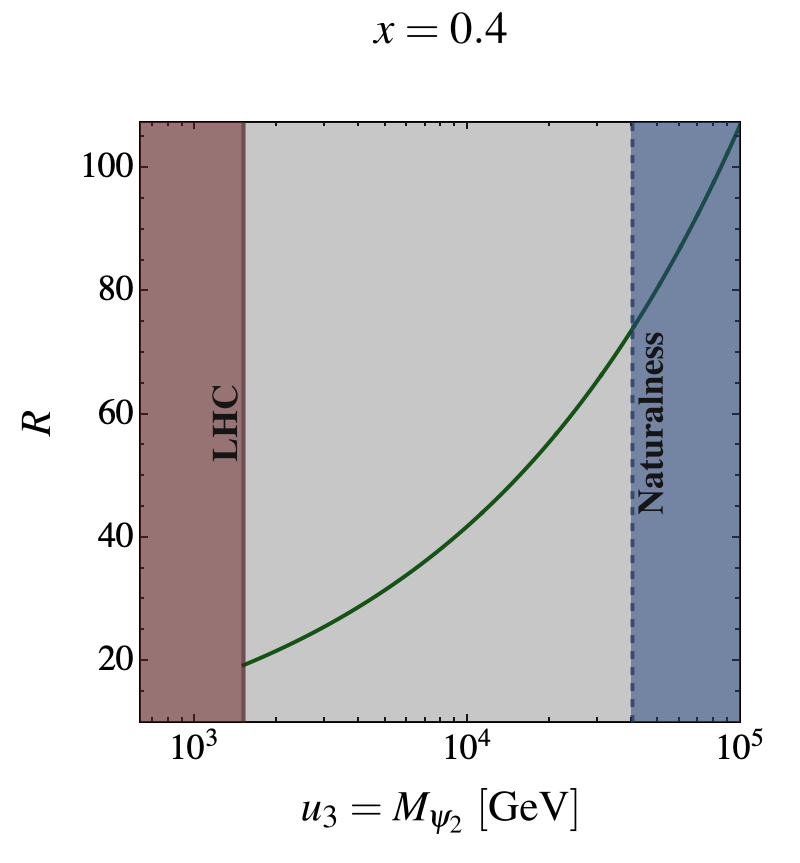}

    \includegraphics[width=0.3\textwidth]{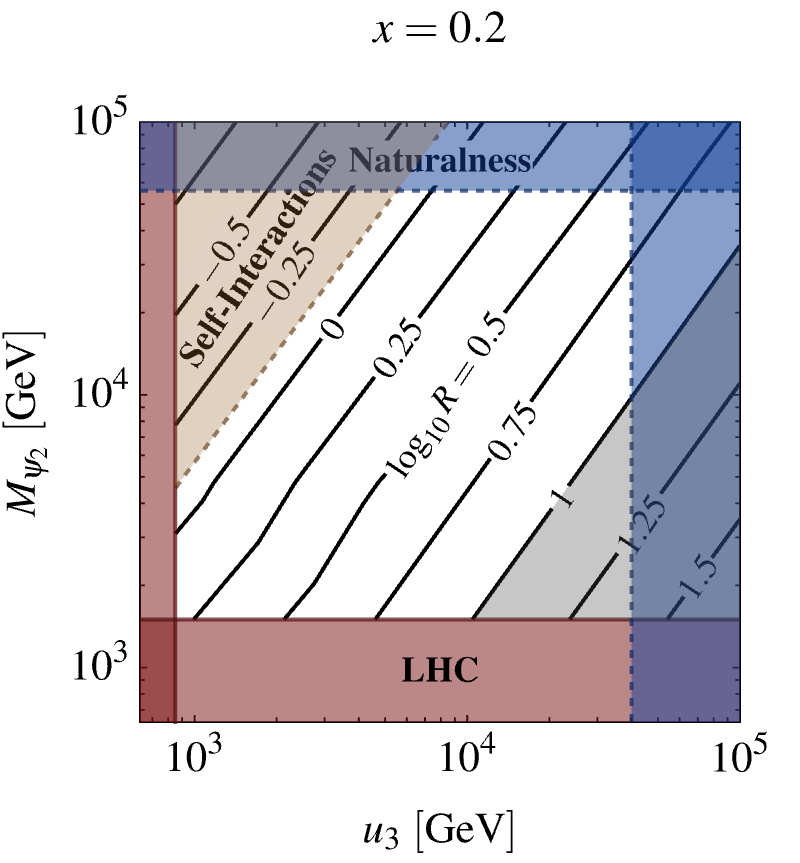}
    \hfill
    \includegraphics[width=0.3\textwidth]{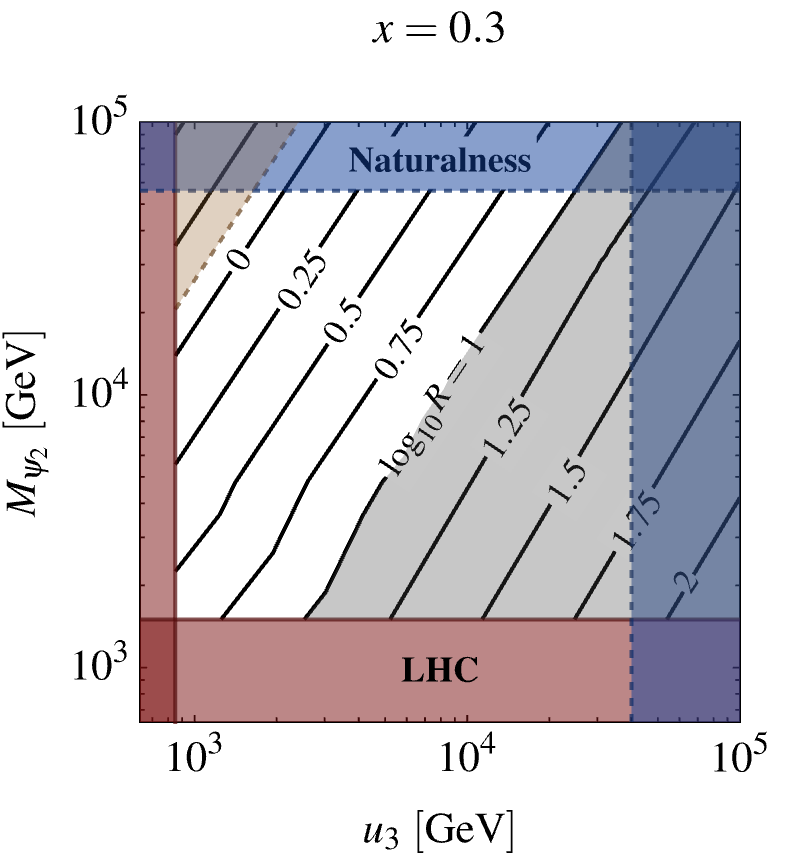}
    \hfill
    \includegraphics[width=0.3\textwidth]{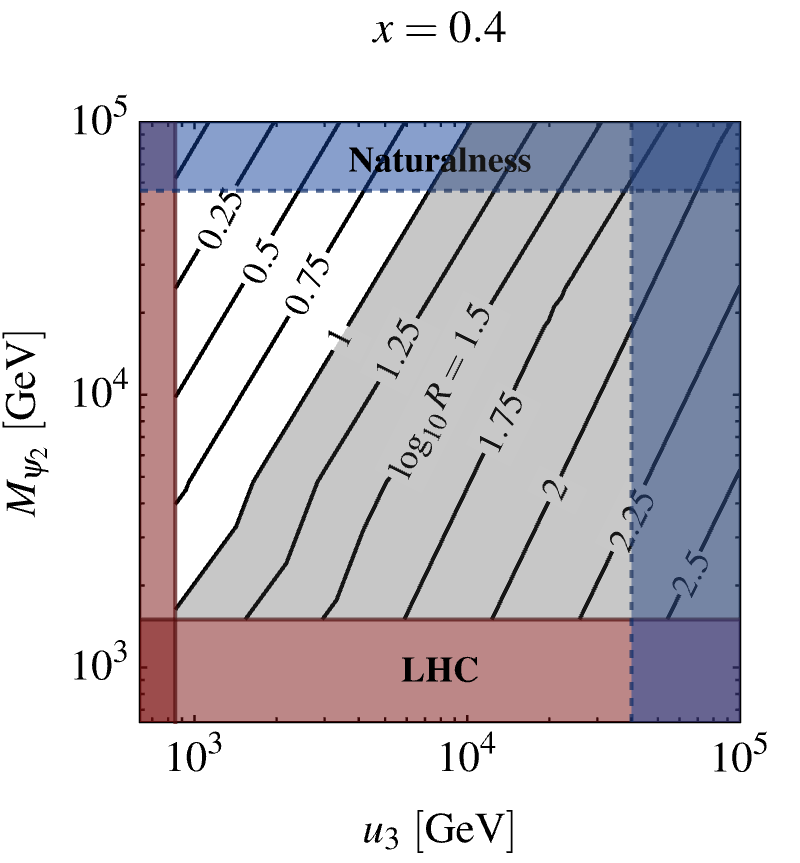}
    
    \caption{Ratios of the visible and dark confinement scales, $R=\Lambda_D/\Lambda_{\text{\tiny{QCD}}}$, for different choices of $x$. The top row shows results for the special case $u_3=M_{\psi_2}$, while the bottom row shows contours of  $R$ in the ($u_3,M_{\psi_2})$ plane. The blue, red and brown shaded regions correspond to the naturalness, collider, and self-interacting DM bounds, respectively, as discussed in the main text. Grey shaded regions denote the parameter space where the ratio of confinement scales exceeds $10$.}
    \label{fig:xPlots}
\end{figure*}

In this appendix, we detail the results obtained by varying the value of $x$ in Eq.~(\ref{eqn:QCDpartitionxdefinition}). Essentially, this parameter encodes how similar the QCD and dark QCD couplings are at the scale $u_3$. The choice $x=0.1$ leads to the highest degree of similarity and is used in our results in the main text. Naturally, different choices of $x$ lead to different results. Figure~\ref{fig:xPlots} shows the results for the ratio of confinement scales for $x=0.2$, $x=0.3$, and $x=0.4$. 

Parameter regions where the ratio $R$ is less than ten are considered the most favourable and shown as unshaded regions in the plots. Obviously, $R < 10$ is not a hard bound, merely indicative, and there may well be ADM scenarios where higher $R$ values are required. Note that the parameter region where $R<10$ decreases as $x$ increases and for $x\gtrsim 0.4$ we find no parameter region where $R < 10$.

\end{document}